\documentclass[conference,compsoc]{IEEEtran}

\usepackage{caption}
\usepackage{colortbl}
\usepackage{xcolor}
\usepackage{pgfplots}
\usepackage{multirow}
\usepackage[T1]{fontenc}
\usepackage{import}
\usepackage{url}
\usepackage{commath}
\usepackage{booktabs}  
\usepackage{amsfonts}     
\usepackage{nicefrac}     
\usepackage{microtype} 
\usepackage{multirow}
\usepackage{graphicx}
\usepackage{xcolor}
\usepackage{subcaption}
\usepackage{array}
\usepackage[linesnumbered,ruled,vlined]{algorithm2e}
\usepackage{hyperref}
\usepackage{flushend}
\usepackage{adjustbox}
\usepackage{paralist}
\usepackage{eso-pic}

\pgfplotsset{compat=1.18}

\definecolor{greenbar}{RGB}{102, 224, 102}
\definecolor{redbar}{RGB}{255, 130, 133}

\newcommand{\gradientbar}[1]{%
    \begin{tikzpicture}
        \def\barwidth{1.0} 
        \def\ballposition{#1} 

        \draw[black, thick] (0,0) rectangle (\barwidth,0.2);
        \fill[left color=greenbar, right color=redbar] (0,0) rectangle (\barwidth,0.2);

        \pgfmathparse{
            ifthenelse(\ballposition<1, 0.099,
            ifthenelse(\ballposition>99, 0.98, \ballposition/100*\barwidth + 0.05))
        }
        \let\ballx\pgfmathresult
        
        \draw[black] (\ballx, 0.105) circle (0.085);
        \fill[white] (\ballx, 0.105) circle (0.085);
        
        \node at (\barwidth + 0.4, 0.105) {\ballposition\%}; 
    \end{tikzpicture}%
}

\def\name{MAYA\xspace}

\SetKwInput{KwInput}{Input}             
\SetKwInput{KwOutput}{Output}  
\newcommand{\rev}[1]{#1}

\AddToShipoutPictureBG{
  \AtPageUpperLeft{
    \raisebox{-2\baselineskip}{\makebox[\paperwidth]{\begin{minipage}{21cm}\centering
      Paper accepted at the 47th IEEE Symposium on Security and Privacy (S\&P 2026)
    \end{minipage}}}
  }
}

\begin{document}

\title{MAYA: Addressing Inconsistencies in Generative Password Guessing \\ through a Unified Benchmark}

\author{
\IEEEauthorblockN{William Corrias}
\IEEEauthorblockA{Sapienza University of Rome\\
Rome, Italy\\
corrias@di.uniroma1.it}
\and
\IEEEauthorblockN{Fabio De Gaspari}
\IEEEauthorblockA{Sapienza University of Rome\\
Rome, Italy\\
degaspari@di.uniroma1.it}
\and
\IEEEauthorblockN{Dorjan Hitaj}
\IEEEauthorblockA{Sapienza University of Rome\\
Rome, Italy\\
hitaj.d@di.uniroma1.it}
\and
\IEEEauthorblockN{Luigi V. Mancini}
\IEEEauthorblockA{Sapienza University of Rome\\
Rome, Italy\\
mancini@di.uniroma1.it}
}

\maketitle
\begin{abstract}
Recent advances in generative models have led to their application in password guessing, with the aim of replicating the complexity, structure, and patterns of human-created passwords. Despite their potential, inconsistencies and inadequate evaluation methodologies in prior research have hindered meaningful comparisons and a comprehensive, unbiased understanding of their capabilities. This paper introduces \name, a unified, customizable, plug-and-play benchmarking framework designed to facilitate the systematic characterization and benchmarking of \rev{deep} generative password-guessing models in the context of trawling attacks. Using \name, we conduct a comprehensive assessment of six state-of-the-art \rev{DL-based models}, which we re-implemented and adapted to ensure standardization, and two \rev{traditional ML-based approaches.} Our evaluation spans eight real-world password datasets and covers an exhaustive set of advanced testing scenarios, totaling over $15,\!000$ compute hours. Our findings indicate that these models effectively capture different aspects of human password distribution and exhibit strong generalization capabilities. However, their effectiveness varies significantly with long and complex passwords. Through our evaluation, \rev{DL-based autoregressive models} consistently outperform other \rev{deep learning approaches}, demonstrating unique capabilities in generating accurate and complex guesses; \rev{meanwhile, ML-based approaches remain surprisingly highly competitive in many scenarios}. Moreover, the diverse password distributions learned by the models enable a multi-model attack that outperforms the best individual model \rev{by an average of $\sim 7$ percentage points}. By releasing \name, we aim to foster further research, providing the community with a new tool to consistently and reliably benchmark generative password-guessing models. Our framework is publicly available at~\url{https://github.com/williamcorrias/MAYA-Password-Benchmarking.git}.
\end{abstract}
\section{\textbf{Introduction}}\label{sec:introduction}
Despite the rise of increasingly secure authentication methods, traditional passwords remain the most widely used mechanism due to their usability and familiarity~\cite{bonneau2012quest,lyastani2020fido2}. However, users often utilize and reuse simplistic passwords across services, making them vulnerable to a variety of attacks~\cite{dell2010password, 6234435, pearman2017let, golla2018site, das2014tangled}. As a result, password security has long been a focus of research. 
Brute-force and dictionary attacks remain the most prevalent password-cracking techniques, primarily exploiting weaknesses such as the use of weak or reused passwords. 
Tools like John The Ripper (JTR)~\cite{jtr} and Hashcat~\cite{hashcat} enhance these techniques by applying transformation rules to dictionaries of passwords, generating variations based on common patterns. Yet, effective transformation rules are complex to design and require extensive contextual knowledge and manual labor. 

Beyond these limitations, traditional Machine Learning password guessing techniques, such as Probabilistic Context-Free Grammars (PCFGs)~\cite{weir2009password, kelley2012guess, cheng2021improved, xu2021chunk, han2020transpcfg} and Markov Models~\cite{ma2014study, durmuth2015omen, guo2021dynamic, xie2022wordmarkov}, generate passwords based on statistical likelihood derived from real-world observed data. While these methods have demonstrated considerable effectiveness, they are fundamentally constrained by structural assumptions. Specifically, Markov Models, which rely on fixed-length n-grams, are inherently limited when modeling long-range dependencies and complex patterns. Likewise, PCFGs impose rigid template structures that constrain the diversity and flexibility of the generated password space.  

In recent years, research has increasingly focused on harnessing advancements in \rev{deep} generative models to enhance password guessing techniques~\cite{passgan, plr-gan, passflow, passgpt, fla, vaepass, rfguess, su2024pagpassgptpatternguidedpassword}. Unlike traditional approaches, these models aim to learn and replicate the complexity, structure, and patterns of human-created passwords without relying on prior assumptions. Nevertheless, the current literature suffers from inconsistencies, inadequate evaluation practices, and a lack of rigorous, standardized characterization of models' behaviors. Proposed approaches are often evaluated using inconsistent methodologies across studies and are tested in non-uniform, often simplistic settings, hindering meaningful comparisons and limiting a full understanding of their capabilities. As a result, critical aspects remain underexplored: how models behave across different settings, what types of passwords they tend to generate, how human-like their guesses are, whether they genuinely capture the underlying complexities of human password behavior, and where potential blind spots may lie. These limitations underscore a broader absence of systematic model characterization.

To address these gaps, we introduce \name 
\footnote{The name "MAYA" is inspired by A. Schopenhauer's philosophical concept of "the veil of Maya", representing an illusion that obscures true reality. Similarly, our framework aims to unveil the potential of password-guessing generative models, which has remained obscured until now.}
, a unified and customizable plug-and-play benchmarking framework. While \name is specifically designed to support fair comparison and in-depth characterization and evaluation of \rev{DL-based} generative password-guessing models in trawling attack scenarios, it builds upon a general, standardized, and rigorous evaluation methodology that can be applied across a wide range of guessing attacks, regardless of the underlying guessing approach. Using \name, we systematically evaluate \rev{eight generative models} for password-guessing, \rev{comprising two ML-based and six DL-based approaches. Each deep generative model was} re-implemented with standardized data preprocessing, dependencies, and configurations. We focus on trawling password attacks, as they represent the most general and widely applicable setting. Our experimental setup spans eight real-world password datasets for training and testing, and covers a wide range of evaluation scenarios, totaling over $15,\!000$ hours of computation. 

Our key findings include: (a) Increasing password length yields diminishing returns in reducing guessability; similarly, increasing the number of generated passwords offers reducing improvements in successful guesses, as models tend to exhaust easily guessable passwords early in the generation process. (b) Rule-based tools \rev{remain} competitive \rev{only} on smaller \rev{datasets, whereas} traditional machine learning models \rev{perform consistently well, particularly} on challenging ones. However, the best-performing \rev{deep learning} model, on average, outperforms both. (c) Generative models do not always require large datasets to effectively model data distribution. While some \rev{approaches} benefit from larger datasets, others exhibit minimal improvement or even a decrease in performance as the training data increases. (d) Models can successfully guess passwords even when trained on datasets from different communities and/or cultures compared to the test set, highlighting strong generalization capabilities and suggesting the existence of common structures underlying human-created passwords across disparate groups. (e) While all models demonstrate strong capabilities in guessing common and simple passwords and a decrease in performance for rarer passwords, \rev{autoregressive DL-based models and traditional ML-based approaches} are the only ones that remain effective in guessing longer passwords and more complex patterns. (f) Different \rev{models} learn generation functions with different probability distributions over the codomain, enabling their combination into a multi-model attack that outperforms individual models. (g) Generative models effectively capture various aspects of human-created passwords, generating high-quality and diverse passwords while minimizing mode failures. However, some models struggle to accurately replicate the length distribution. 
\\\\This paper makes the following contributions:
\begin{itemize}
    \item We analyze eight real-world leaked password datasets, providing a detailed characterization of each and examining the impact of factors such as dataset size, geographic origin, linguistic and cultural background, and temporal span on the resulting password distributions.
    \item We propose a rigorous and standardized evaluation methodology for trawling password-guessing attacks, addressing key methodological gaps in prior work. Our proposal enables fair comparisons, a rigorous empirical evaluation, and a comprehensive characterization of each approach.
    \item We develop \name, a fully customizable, plug-and-play framework for evaluating generative password-guessing models in trawling attack scenarios. To support reproducibility and encourage further research, we publicly release our code and data at~\url{https://github.com/williamcorrias/MAYA-Password-Benchmarking.git}.
        \item Leveraging \name, we characterize and benchmark \rev{eight} state-of-the-art generative models across eight datasets and a diverse set of testing scenarios, addressing seven key research questions. Our evaluation spans over $15,\!000$ compute hours.
    \item We provide insights to guide future research in enhancing model password-guessing capabilities and integrating them into other password-related domains.
\end{itemize}
\section{\textbf{Motivation}}\label{sec:motivation}
The current body of research on generative password-guessing models suffers from several notable limitations. First, existing evaluation methodologies are inconsistent across studies, hindering meaningful comparisons. Second, evaluations often lack methodological rigor, typically relying on overly simplified scenarios that fail to provide a comprehensive or unbiased assessment of model performance. Third, there is a lack of systematic characterization, which restricts our understanding of what these models learn and how they behave under different conditions. These limitations motivate the need for a unified and comprehensive benchmarking framework that enables fair comparisons, subjects models to advanced and realistic scenarios, and offers a deeper insight into their underlying capabilities.

\subsection{\textbf{Lack of Consistency}}
Each existing approach adopts its own evaluation methodology, as models are neither trained nor assessed on the same data and under identical settings. These methodological inconsistencies encompass various factors, including data preprocessing algorithms, file encoding, vocabulary, maximum password length, the number of generated passwords, and the size of the training and testing datasets. \rev{For example, (1) PassGAN~\cite{passgan} compares models with different numbers of generated passwords; (2) VGPT2~\cite{vgpt2} and FLA~\cite{fla} define custom alphabets with ad-hoc special characters, whereas~\cite{passgpt, passgan, plr-gan, passflow} rely on the complete UTF-8 set; (3) PassFlow~\cite{passflow} is trained on a maximum of $300.000$ passwords and PassGPT~\cite{passgpt} is trained only on unique passwords, while ~\cite{fla, passgan, plr-gan, vgpt2} adopt the standard 80-20 split without additional sampling; (4) FLA~\cite{fla} filters passwords based on a minimum length, whereas~\cite{vgpt2, passgpt, passgan, plr-gan, passflow} utilize a maximum length threshold (VGPT2 using 12, and the others 10).} As a result, fair comparisons across studies are challenging. For instance, differing data preprocessing methods lead to distinct training and testing distributions, making direct comparison challenging. Likewise, evaluating models with different maximum password lengths introduces inherent bias, as shorter passwords are generally easier to guess. Analogous considerations apply to other settings as well. Such inconsistencies highlight the necessity for a standardized evaluation methodology to serve as a foundation for future research. 

\subsection{\textbf{Lack of Rigorousness}}
Existing research often suffers from an insufficiently rigorous evaluation. Models are typically evaluated using overly simplistic metrics within restricted and similarly simplistic scenarios, such as the percentage of guessed passwords or the number of unique passwords generated \rev{~\cite{passgan, vgpt2, passflow}}. \rev{Moreover, part of the literature ~\cite{passgan, passflow}} relies only on widely used datasets like RockYou or LinkedIn. Consequently, the current literature offers an incomplete evaluation biased toward these simplistic settings, underscoring the need for more robust methodology incorporating diverse datasets, varied scenarios, and more complex settings that reflect a broader spectrum of real-world contexts.

\subsection{\textbf{Lack of Characterization}}
The methodological issues outlined above lead to a broader problem: the lack of a systematic characterization of password-guessing models. Beyond aggregate performance metrics, current research fails to offer in-depth insights into crucial aspects of generative models, such as the human-likeness of generated passwords, the structural properties of real passwords these models capture, the types of passwords they generate (and those they fail to generate), the distinct distributions that different models learn, and how model behavior varies across different experimental conditions. Such insights are essential for a comprehensive understanding of these models, their true capabilities, and potential applications beyond password guessing.

\subsection{\textbf{Why Trawling Attacks}}
Password attacks are typically categorized based on the adversary’s scope: targeted and trawling. Targeted attacks aim to compromise specific user accounts by leveraging personal identifiable information (PII), and have gained increasing attention in recent years due to their growing effectiveness~\cite{li2016study, rfguess, pal2019beyond, wang2023pass2edit, he2022passtrans, xu2023improving}. In contrast, trawling attacks seek to guess as many user passwords as possible within a dataset, without targeting any specific individual. This attack model has long been the focus of password security research, with numerous techniques developed over time, including rule-based systems, PCFGs, Markov models, traditional neural networks, and deep generative models. In this work, we focus exclusively on trawling attacks, as they represent the most general and widely applicable class of password-guessing attacks, providing a broad evaluation framework. \rev{Targeted password-guessing attacks require different evaluation scenarios and datasets, and do not properly align with our RQs. Nevertheless, we plan to extend MAYA in future work to include them.} Moreover, since targeted attacks can be viewed as trawling attacks conditioned on PII, our evaluation methodology can effectively support both lines of research.
\section{\textbf{\name}}\label{sec:framework}
This section presents \name, our unified and plug-and-play benchmarking framework tailored for \rev{deep generative} password-guessing models in the context of trawling attacks. The framework offers an intuitive environment for training and testing models with minimal setup, enabling the research community to move in a unified direction. It also includes a comprehensive set of experimental settings, providing thorough benchmarking and detailed characterization across multiple key metrics. \name features a highly modular and easily extendable architecture that enables the integration of new models and customized testing scenarios to support future research. While \name focuses on \rev{DL-based approaches}, its underlying evaluation methodology (Section~\ref{sec:framework-approach}) provides a standardized testing environment that is applicable to all types of trawling password-guessing methods, ranging from traditional techniques to the latest approaches. \name currently implements six state-of-the-art \rev{deep generative} password-guessing models \rev{and two traditional ML-based approaches} (Section~\ref{sec:framework-models}), eight real-world password datasets (Section~\ref{sec:datasets}), and a comprehensive set of testing scenarios aimed at answering seven key research questions (Section~\ref{sec:framework-scenarios}).

\subsection{\textbf{Methodology}}\label{sec:framework-approach}
\name addresses the methodological shortcomings identified in Section~\ref{sec:motivation} by introducing a standardized and rigorous evaluation methodology. This methodology enables fair comparisons across approaches and supports thorough benchmarking and detailed characterization of trawling password-guessing methods.

\subsubsection{\textbf{Standardized Data Preprocessing and Settings}}\label{sec:framework-approach-dp}
We propose the following procedure to standardize data preprocessing: (1) open and read datasets using UTF-8 encoding while ignoring errors, (2) remove passwords that exceed the specified maximum length, contain non-ASCII characters, or include characters that are not in the provided vocabulary. (3) split the dataset following the standard \(80\%\) training and \(20\%\) testing ratios, (4) remove duplicates from the testing dataset, and (5) remove from the training dataset any overlap with the testing dataset. 
This approach ensures wide language compatibility through UTF-8 encoding, avoids double-counting by eliminating duplicates, and provides a fixed, consistent testing set across experiments. We note that existing works typically define the training set first, and then derive the testing set by removing any overlapping samples. This approach is undesirable, as varying the training set leads to changes in the testing set as well, thereby limiting the comparability across experiments.
We further define our vocabulary to include all uppercase and lowercase letters, digits, and the following widely-accepted symbols: \verb`~!@#$%^&*(),.<>/?'"{}[]\|-_=+;: ``

\subsubsection{\textbf{Advanced Evaluation Scenarios}}\label{sec:framework-approach-ats}
We have designed a set of advanced and realistic evaluation scenarios to accurately and comprehensively assess the models’ capabilities and limitations, with the goal of providing a complete and nuanced understanding of their performance. Each scenario has been envisioned to address one of the seven research questions outlined in Section~\ref{sec:framework-scenarios}.

\subsection{\textbf{Models}}\label{sec:framework-models}

\begin{table}[t]
    \centering
    \caption{\rev{Categorization of the selected approaches.}}
    \label{tab:categorization}
    \renewcommand{\arraystretch}{1.15}
    \begin{adjustbox}{width=0.75\columnwidth, center}
    \begin{tabular}{l *{5}{l}}
        \toprule
        \textbf{\rev{Model}} & \textbf{\rev{Trad. ML}} & \textbf{\rev{DL-Based}} & \textbf{\rev{Latent-Based}} & \textbf{\rev{Autoregressive}} \\
        \midrule
        FLA & X & \checkmark & X & \checkmark \\ 
        OMEN & \checkmark & X & X & \checkmark \\
        PassFlow & X & \checkmark & \checkmark & X \\ 
        PassGAN & X & \checkmark & \checkmark & X \\ 
        PassGPT & X & \checkmark & X & \checkmark \\ 
        PCFG & \checkmark & X & X & X \\
        PLR-GAN & X & \checkmark & \checkmark & X \\ 
        VGPT2 & X & \checkmark & \checkmark & X \\
        \bottomrule
    \end{tabular}
    \end{adjustbox}
\end{table}

\rev{Table~\ref {tab:categorization} shows the selected models and their respective categories used throughout the paper for reference. We selected eight generative models: two ML-based and six DL-based. Among them, four are latent-based approaches and three are autoregressive. We consider a model latent-based if it directly operates over a latent space $z$~\cite{hu2023complexity}. 

Regarding DL-based models, we aimed to cover diverse architectural families}: (1) FLA~\cite{fla}, \rev{an autoregressive approach} based on an LSTM, (2) PassGAN~\cite{passgan}, and (3) PLR-GAN~\cite{plr-gan}, both GAN-based \rev{and thus latent-based}, (4) PassFlow~\cite{passflow}, \rev{a latent-based model built on a flow architecture}, (5) VGPT2~\cite{vgpt2}, \rev{a latent-based approach combining} a VAE with GPT2-derived encoder and decoder blocks, and (6) PassGPT~\cite{passgpt}, an autoregressive transformer model based on the GPT2 architecture. \rev{Our experiments also include two traditional ML-based password-guessing approaches: PCFG~\cite{weir2009password} relying on Probabilistic-Context-Free-Grammars, and OMEN~\cite{durmuth2015omen} based on Markov Chains. This allows for a comprehensive comparative analysis between ML-based and DL-based methods, offering a complete overview and characterization of each approach.}

\rev{Each DL-based model was fully or partially re-implemented to (1) standardize dependencies, and (2) enable a unified comparison framework. Specifically, we ported them to PyTorch 2.6.0 and adapted them to \name's interface. These changes provide a common playground for fair comparisons and facilitate future research to work within the same testbed.} We validated the accuracy of our implementations by comparing our results with those reported in the original papers, \rev{observing only negligible variations falling within the margin of error. Whenever possible, we also contacted the original authors for additional validation.} We provide further implementation details in Appendix~\ref{sec:appendix-models}.

\subsubsection*{\textbf{FLA}}
FLA~\cite{fla} (Fast, Lean, and Accurate) was the first approach to apply neural networks to the password-guessing task. It is the only approach we selected based on a recurrent neural network, specifically an LSTM, allowing us to examine how \rev{autoregressive} models perform compared to others, more recent generative architectures.

\subsubsection*{\textbf{PassGAN}}
PassGAN~\cite{passgan} is based on a Generative Adversarial Network (GAN) architecture, which, unlike other designs, follows an adversarial training approach. As they are implicit models, GANs learn to generate data by capturing the underlying distribution of the training set without explicitly defining a probability distribution, offering a unique perspective in password generation.

\subsubsection*{\textbf{PLR-GAN}}
PLR-GAN~\cite{plr-gan} is an enhanced version of PassGAN and represents the state-of-the-art for GAN-based methods. PLR-GAN adopts a Dynamic Password Guessing (DPG) strategy, which allows the model to adapt its guesses based on the distribution of successfully guessed passwords, increasing the likelihood of generating relevant guesses. 

\subsubsection*{\textbf{PassFlow}}
Passflow~\cite{passflow} represents the first and only attempt to integrate flow-based generative models into the field of password guessing. Flow networks~\cite{papamakarios2021normalizing} offer an explicit latent space and an invertible mapping between a data point and its latent representation, enabling complex operations such as interpolation and exact latent variable inference. PassFlow adopts and further enhances DPG by integrating Gaussian Smoothing in the generation process, reducing the likelihood of generating duplicated passwords while maintaining the benefits of DPG.

\subsubsection*{\textbf{VGPT2}}
VGPT2~\cite{vgpt2} combines a Variational Autoencoder (VAE) with an encoder-decoder architecture based on GPT2. The VAE provides an explicit representation of the latent space, while GPT2 excels at processing sequential information and capturing long-term dependencies, making it a unique approach to analyze.

\subsubsection*{\textbf{PassGPT}}
PassGPT~\cite{passgpt} proposes a GPT-2-based language model for password generation. Similarly to FLA, PassGPT employs an \rev{autoregressive} generation process. However, GPT-2 is built upon an attention mechanism, which allows it to capture long-range dependencies more effectively.

\rev{\subsubsection*{\textbf{PCFG}}
PCFG~\cite{weir2009password} learns a grammar -i.e., a set of rules describing password structures- from a dictionary of passwords. Given a structure (e.g., 4 letters followed by 4 digits), PCFG fills it using terminals. Thus, PCFG explicitly learns both the password patterns and the probabilities of each terminal within a pattern. The final probability of a password is calculated by multiplying the probability of the structure by the probabilities of the selected terminals. Passwords are generated in order of decreasing probability.}

\rev{\subsubsection*{\textbf{OMEN}}
OMEN~\cite{durmuth2015omen} is a variant of the Markov n-gram model that generates passwords in descending probabilistic order, achieving higher performance than its traditional counterpart. Unlike PCFG, OMEN does not explicitly learn password patterns; instead, these are fixed and determined by the chosen n-gram window. What OMEN actually learns are the probabilities of character transitions within each window.}

\subsection{\textbf{Research Questions}}\label{sec:framework-scenarios}
This section presents the research questions and scenarios that guided the design of our evaluation, enabling the comprehensive characterization of model behavior and further comparison of their performance.

\subsubsection*{\textbf{RQ1 - How Do Different Settings Influence Models Performance?}} Generative password-guessing models have two primary settings: the maximum password length and the number of generated passwords. This RQ explores the impact of these two factors on guessing performance. We test three different maximum lengths (8, 10, and 12) and eleven generation quantities, from $10^6$ to $5 \times 10^8$. For each dataset, all models were trained three times—once per length—and evaluated across all generation quantities.

\subsubsection*{\textbf{RQ2 - Are \rev{Deep} Generative Models Truly Better than Traditional Tools?}}
Despite recent advancements in \rev{deep} generative architectures, there is limited evidence indicating whether \rev{DL-based generative} models outperform traditional tools \rev{and ML-based models, with only a} few direct experimental comparisons \rev{conducted}.
We address this gap by conducting a comprehensive evaluation of generative models on 8 different datasets and comparing them to traditional tools such as John the Ripper (JtR) and Hashcat, offering a thorough, direct comparison between \rev{DL-based} generative, \rev{ML-based generative}, and rule-based methods. 

\subsubsection*{\textbf{RQ3 - How Sensitive are Models to Training Dataset Size?}}
In real-world scenarios, attackers often have limited or partial access to leaks. Therefore, evaluating model performance across varying training data subset sizes is essential for gaining a clear understanding of the effectiveness of generative models. This RQ examines the ability of the models to reconstruct the full target data distribution starting from varying portions of the source dataset and execute a successful attack. We explore this capability by training models on up to seven different data subsets on four datasets.

\subsubsection*{\textbf{RQ4 - Can Models Generalize To Different Communities and/or Cultures?}}
A common real-world scenario involves attackers obtaining leaked passwords from a distribution different from their intended target, often due to cultural or community differences. This RQ investigates whether generative models can effectively generalize to unseen password distributions. We assess this capability through extensive cross-dataset analysis, testing the models on datasets distinct from those used during training. Specifically, we examine their ability to guess passwords across (1) different communities (i.e., same language but different online services) and (2) different cultures (i.e., different languages and cultural backgrounds), which, as demonstrated in Section~\ref{sec:datasets-analysis}, significantly impacts password distributions.

\subsubsection*{\textbf{RQ5 - Are Models Limited to Guessing Only Simple and Common Passwords?}}
Simple passwords are easily guessed using traditional tools, whereas rare and complex passwords present a significantly greater challenge. The ability of generative models to arbitrarily sample from different areas of the password distribution offers the potential to guess even rare and complex passwords. We explore this RQ by dividing and categorizing the test dataset based on password frequency and assessing each model's performance on the different subsets. Additionally, we analyze how models perform in guessing passwords of varying lengths and patterns. 

\subsubsection*{\textbf{RQ6 - To What Extent Do the Distributions Learned by Different Models Align? Can We Combine Models to Maximize Effectiveness?}}\label{ref:rq6}
Generative models are trained to generate data that matches the target distribution, learning this distribution either implicitly or explicitly. However, it remains unclear to what extent distributions learned by different models align, as various factors influence the learning process. If the distributions are not fully aligned, a multi-model attack could enhance guessing capabilities. 

\paragraph{\textbf{RQ6.1 - To What Extent Do the Distributions Learned by Different Models Align?}}
We explore the first part of this RQ using two metrics computed across all model pairs $(M_1,M_2)$: Jaccard Index and Mergeability Index. Jaccard Index measures the ratio between the intersection and union of the passwords generated by each model, \rev{providing a metric of diversity between $M_1$ and $M_2$ passwords}.
Additionally, we introduce the Mergeability Index, which quantifies the ratio between the marginal gain achieved by combining the outputs of $M_1$ and $M_2$ over the best performance between $M_1$ and $M_2$ (see Eq.~\ref{equation:mergeability}). This metric assesses the weighted improvement gained by combining $M_1$ and $M_2$, compared to using only the best-performing model.

\paragraph{\textbf{RQ6.2 - Can We Combine Models to Maximize Effectiveness?}}
We follow an iterative elimination approach to gain insights into whether and to what extent combining multiple models enhances guessing effectiveness. We begin by combining passwords generated by all models and progressively remove those generated by the least effective model in terms of additional matched passwords. This process allows us to iteratively refine and identify effective combinations of models for any desired number of models to combine.

\subsubsection*{\textbf{RQ7 - Do Models Truly Capture the Characteristics of Human-Like Passwords?}}
Despite ample research, it remains unclear to what extent the distribution learned by generative models aligns with that of human-created passwords. Evaluation metrics such as guess percentage are inherently limited, as they only measure whether generated passwords match the test data, without offering insights into the distribution of non-matching guesses. A major challenge in answering this RQ is how to quantify the distance between generative models' password distribution and the overall distribution of human-like passwords. In related fields, metrics such as Fréchet Inception Distance (FID)~\cite{FID} and Inception Score (IS)~\cite{IS} are commonly used to assess the generated data~\cite{borji2022pros}. However, these metrics rely on pre-trained classifiers and are unsuitable for the password domain. We address this gap by identifying and adopting four alternative metrics: (1) CNN Divergence~\cite{cnndivergence}, (2) IMD~\cite{imd}, (3) $\alpha$-Precision $\beta$-Recall Authenticity~\cite{precisionrecall}, and (4) MTopDiv~\cite{mtopdiv}. These metrics were carefully selected for their ability to capture different aspects of the considered distribution. Appendix~\ref{ref:appendix_rqs} provides a detailed explanation of each metric. We computed these metrics for each model across all datasets and averaged the results, offering a comprehensive assessment of the human-likeness of the generated passwords. Additionally, we complement this analysis by examining the length distribution of the generated passwords and comparing it to that of real passwords.
\section{\textbf{Datasets}}\label{sec:datasets}
Our framework incorporates eight real-world leaked password datasets, carefully chosen to ensure diversity in size, geographic origin, language, time of leakage, and service type. In the following sections, we provide a detailed description of each dataset and perform a statistical analysis to examine the distribution and characteristics of the passwords across these datasets, providing insights that form the basis for our subsequent experiments.

\subsection{\textbf{Datasets Selection}}\label{sec:datasets-selection}

\begin{table}[t!]
    \centering
    \caption{Details of the selected datasets.}
    \label{tab:dataset_table}
    \renewcommand{\arraystretch}{1.15}

    \begin{adjustbox}{width=0.98\columnwidth, center}
    \begin{tabular}{l|l l l l l l |l l|l c}
        \toprule
        \multicolumn{1}{c}{\textbf{}} & \multicolumn{6}{c}{\textbf{General}} & \multicolumn{2}{c}{\textbf{\rev{Filtering}}} & \multicolumn{2}{c}{\textbf{\rev{Policy}}} \\
        \cmidrule(r){1-7} \cmidrule(lr){8-9} \cmidrule(l){10-11}
        {\textbf{Dataset}} & {\textbf{\#Password}} & {\textbf{\#Unique}} & {\textbf{Loc}} & {\textbf{Lang}} & {\textbf{Year}} &{\textbf{Service}} &{\textbf{\rev{\#Rem}}}  &{\textbf{\rev{\%Rem}}} & {\textbf{\rev{Char}}} & {\textbf{\rev{Len}}} \\
        \midrule
        Rockyou~\cite{rockyou} & 32.600.024 & 14.311.994 & USA & EN & 2009 & Gaming & 17.788 & 0.05\% & NONE & $\ge$ 5 \\
        Linkedin~\cite{linkedin} & 60.650.662 & 60.591.405 & Global & EN & 2012 & Social & 0 & 0.00\% & NONE & $\ge$ 6 \\
        Mail.ru~\cite{mailru} & 3.723.472 & 2.260.454 & RU & RU & 2014 & Mail & 0 & 0.00\% & NONE & $\ge$ 5 \\
        000webhost~\cite{000webhost} & 15.269.739 & 10.587.879 & USA & EN & 2015 & Forum & 18.742 & 0.12\% & 2 Classes & $\ge$ 6 \\
        Taobao~\cite{taobao} & 7.492.035 & 6.165.957 & CHN & ZH & 2012 & Ecomm & 5 & 0.00\% & NONE & $\ge$ 6 \\
        Gmail~\cite{gmail} & 4.912.520 & 3.122.573 & RU & RU & 2014 & Mail & 10.453 & 0.21\% & NONE & - \\
        AshleyMad.~\cite{ashleymadison} & 375.846 & 375.738 & CA & EN & 2015 & Social & 0 & 0.00\% & NONE & - \\
        Libero~\cite{libero} & 667.680 & 418.400 & IT & IT & 2016 & Mail & 331 & 0.05\% & NONE & $\ge$ 6 \\
        \bottomrule
    \end{tabular}
    \end{adjustbox}
\end{table}

Table~\ref{tab:dataset_table} presents the datasets selected for this study. \rev{For ethical reasons, we avoided newer and larger datasets sold in unverified forums or websites and focused on publicly available ones used in previous works.} To ensure that the selected datasets provide generalizable insights, the following criteria were considered in the selection process:

\subsubsection*{\textbf{Dataset Size}} We included datasets of varying sizes to evaluate model performance under different dimensions. The collection ranges from small datasets with a few hundred thousand passwords, such as Libero~\cite{libero}, to large-scale datasets with tens of millions of passwords, such as RockYou~\cite{rockyou}, LinkedIn~\cite{linkedin}, and 000webhost~\cite{000webhost}. \rev{We excluded archives comprising aggregations of multiple datasets, as they would introduce biases in generalizability analysis and may overlap with the other datasets used}.

\subsubsection*{\textbf{Geographic Diversity}} The selected datasets vary in their geographical provenance. Previous studies have shown that users from different countries exhibit distinct password creation behaviors, leading to diverse password distributions~\cite{zipf_password_1, analysis_chinese_web_passwords, 9836812, 7298428, wang2016implications}. \rev{This offers an opportunity to assess models across different password patterns}.

\subsubsection*{\textbf{Language and Cultural Background}} We selected datasets representing diverse languages and \rev{specific cultural contexts. This allows us to study the generalizability between different cultures}, as users from different backgrounds exhibit variations in password choices~\cite{zipf_password_1, alsabah2018your, analysis_chinese_web_passwords, wang2016implications}.

\subsubsection*{\textbf{Temporal Span}} We consider the dataset leak date as an important factor, as password policies and user awareness have evolved over time, with stronger requirements introduced in recent years~\cite{205136, 10.1145/1940976.1940994, furnell2011assessing}. Consequently, we expect older leaks to contain weaker passwords, while more recent datasets are expected to contain more secure passwords. Our dataset collection spans from 2009 to 2016. 

\subsubsection*{\textbf{Service Types and Community Background}} 
We selected datasets from a wide range of services and communities, such as social networks, forums, e-commerce sites, dating sites, email services, and gaming platforms. \rev{While some datasets are highly community-specific, they were specifically selected to assess generalizability across niche or very different communities.} The type of service and the nature of the user community can significantly influence password selection strategies, with more sensitive services and close-knit communities often prompting users to adopt different password behaviors~\cite{stobert2018password, abdrabou2022your, jin2024password}. 

\subsection{\textbf{Dataset Analysis}}\label{sec:datasets-analysis}

\begin{table}[t]
    \centering
    \caption{Distribution of password lengths.}
    \label{tab:len_distribution}
    \begin{adjustbox}{width=0.95\columnwidth, center}
    \begin{tabular}{l *{10}{c}} 
        \toprule
        \textbf{Dataset} & {\textbf{1-5}} & {\textbf{6}} & {\textbf{7}} & {\textbf{8}} & {\textbf{9}} & {\textbf{10}} & {\textbf{11}} & {\textbf{12}} & {\textbf{13+}} \\
        \midrule
        rockyou & 4.33 & \textbf{26.06} & 19.30 & 19.99 & 12.12 & 9.06 & 3.56 & 2.10 & 3.49 \\
        linkedin & 0.00 & 8.95 & 11.01 & \textbf{24.60} & 14.47 & 12.54 & 6.56 & 4.19 & 17.68 \\
        mailru & 1.78 & 22.90 & 14.68 & \textbf{23.92} & 11.64 & 8.81 & 5.22 & 4.31 & 6.74 \\
        000webh & 0.06 & 5.71 & 7.94 & \textbf{21.88} & 15.41 & 14.50 & 10.48 & 7.66 & 16.36 \\
        taobao & 0.48 & 12.93 & 13.14 & \textbf{16.90} & 16.51 & 16.10 & 10.77 & 6.80 & 6.39 \\
        gmail & 4.13 & 18.73 & 13.49 & \textbf{28.92} & 13.85 & 13.85 & 3.10 & 1.89 & 2.05 \\
        ashleym & 9.97 & 19.57 & 18.15 & \textbf{24.87} & 13.21 & 9.66 & 2.19 & 1.26 & 1.12 \\
        libero & 0.09 & 15.49 & 11.34 & \textbf{31.30} & 16.34 & 11.32 & 5.54 & 3.83 & 4.76 \\
        \midrule
        Average & 2.60 & 16.29 & 13.63 & \textbf{24.05} & 14.19 & 11.98 & 5.93 & 4.01 & 7.32 \\
        CDF & 2.60 & 18.90 & 32.53 & 56.58 & 70.77 & 82.75 & 88.68 & 92.68 & 100.00 \\
        \bottomrule
    \end{tabular}
    \end{adjustbox}
\end{table}

This section presents the main insights obtained from our statistical analysis of the selected datasets. 

\subsubsection{\textbf{Password Length Analysis}} 
Table~\ref{tab:len_distribution} shows that the most common lengths are between 6 and 10 characters. Taobao, 000webhost, LinkedIn, and Libero enforce a stricter policy (len >= 6), with very few passwords shorter than 6 characters. The Cumulative Distribution Function (CDF) reveals that over $50\%$ of passwords are at most 8 characters long, with 8 being, on average, the most frequently chosen length. Notably, the CDF rapidly increases up to 12 characters, suggesting that passwords longer than 12 characters are rare. Based on these observations, we selected passwords with maximum lengths of 8, 10, and 12 in our experiments. 

\subsubsection{\textbf{Password Patterns Analysis}} 
Based on common password rules and user behaviors, we defined 19 patterns, listed in Table~\ref{tab:regex}, to characterize password patterns across the datasets. For a detailed distribution analysis, see Table~\ref{tab:pattern_distribution} in the Appendix. In most datasets, letter-only passwords are highly prevalent and are almost always exclusively lowercase, whereas uppercase-only or mixed-case passwords appear infrequently, apart from 000webhost, likely influenced by a stricter password policy enforced prior to the breach. In European and American datasets, digit-only passwords are less frequent than letter-only ones, whereas Taobao (from China) shows a higher prevalence of digit-only passwords, consistent with prior studies~\cite{zipf_password_1, alsabah2018your, analysis_chinese_web_passwords}. This is likely due to users being less familiar with the Latin alphabet. Passwords comprised entirely of special characters are rarely used, as they are hard to remember, and users prioritize usability over security. A common pattern across all datasets is the combination of letters and digits. In 000webhost $\approx93\%$ of passwords follow this pattern, suggesting a policy requiring at least two character classes. 
Users often find password policies frustrating~\cite{shay2010encountering, shay2016designing, komanduri2011passwords} and tend to fall into predictable patterns to comply with these requirements, such as appending a trailing digit~\cite{zipf_password_1, wang2016implications}. Our analysis supports this observation, as passwords ending with a digit are highly prevalent, with nearly half of all passwords following this pattern and ‘1’ being widely used as a final character. 

\begin{table}[t]
    \centering
    \caption{Selected patterns and their description.}
    \label{tab:regex}
    \begin{adjustbox}{width=0.95\columnwidth, center}
    \begin{tabular}{ll ll}
        \toprule
        \textbf{ID} & \textbf{Description} & \textbf{ID} & \textbf{Description} \\
        \midrule
        r1 & Letters only. & r10 & Starts letter ends digit. \\
        r2 & Lowercase letters only. & r11 & Starts letter ends special. \\
        r3 & Uppercase letters only. & r12 & Starts digit then only letters. \\
        r4 & Digits only. & r13 & Starts digit ends special. \\
        r5 & Special only. & r14 & Starts and ends with digit. \\
        r6 & Letters and digits. & r15 & Starts special, then only letters. \\
        r7 & Letters and special. & r16 & Starts and ends with special. \\
        r8 & Digits and special. & r17 & Starts special, ends digit. \\
        r9 & Letters, digits, and special. & r18 & Ends with '!'. \\
           &                        & r19 & Ends with '1'. \\
        \bottomrule
    \end{tabular}
    \end{adjustbox}
\end{table}

\begin{figure}[t]
    \centering
    \subfloat[RockYou\label{fig:zipf-rockyou}]{
    \includegraphics[width=0.45\columnwidth]{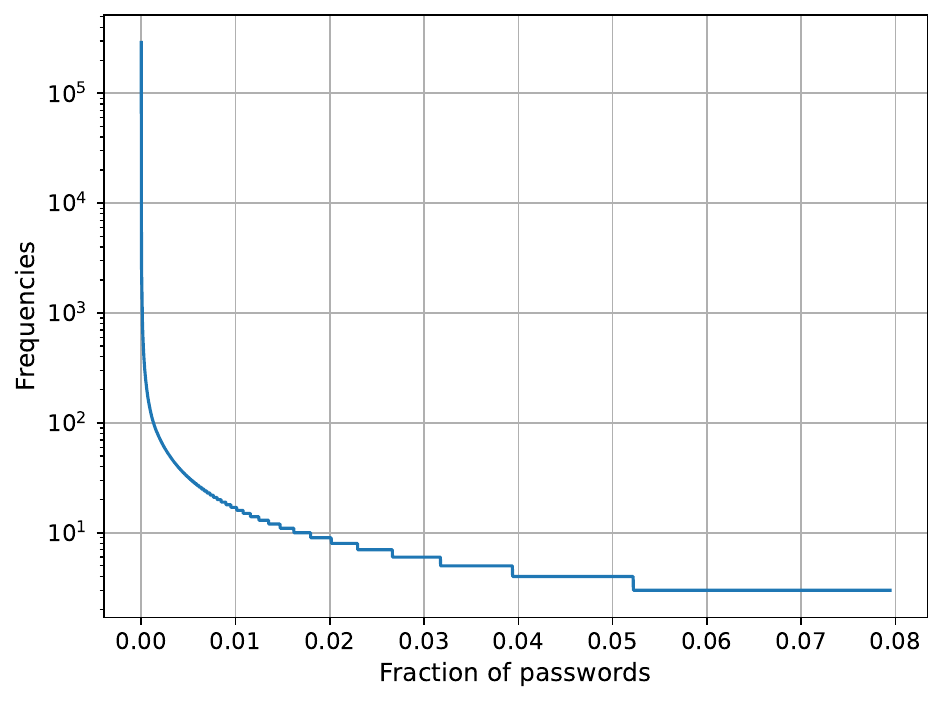}
    }
    \subfloat[000webhost\label{fig:zipf-000webhost}]{
    \includegraphics[width=0.45\columnwidth]{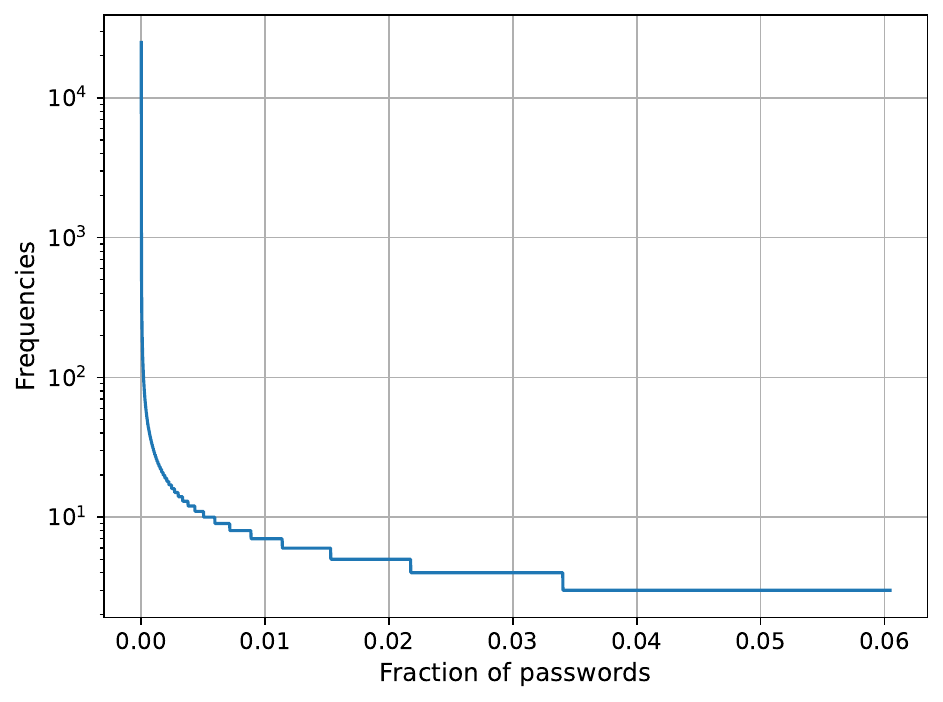}
    }
    \caption{Password frequency distribution. The x-axis is computed over the total number of unique passwords.}
    \label{fig:zipf-distributions}
\end{figure}

\subsubsection{\textbf{Top-10 Passwords Analysis}} 
We analyzed the top 10 passwords in each dataset, finding that common choices like ‘123456’ appear across almost all datasets. Notably, password preferences vary by region: European users favor lowercase-only passwords, Chinese users often choose digit-based passwords resembling words phonetically, and Russian users prefer keyboard patterns. Table~\ref{tab:top10} in the Appendix presents the top 10 most common passwords.

\begin{figure*}[t!]
    \centering
    \subfloat[Max. Password Length 8]{
      \includegraphics[width=0.24\textwidth]{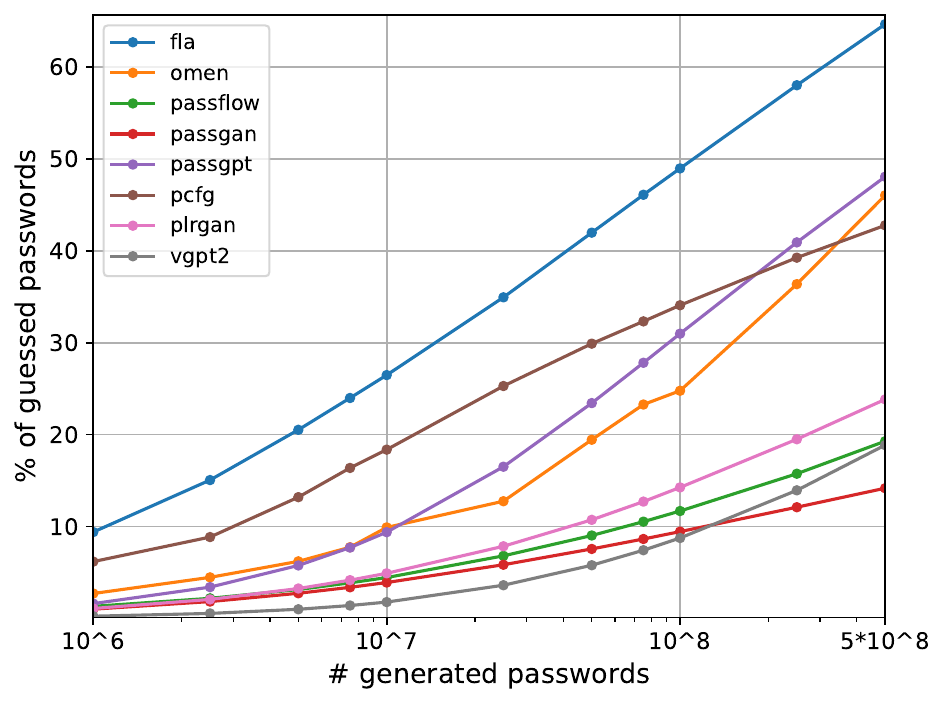}
    }
    \subfloat[Max. Password Length 10]{
      \includegraphics[width=0.24\textwidth]{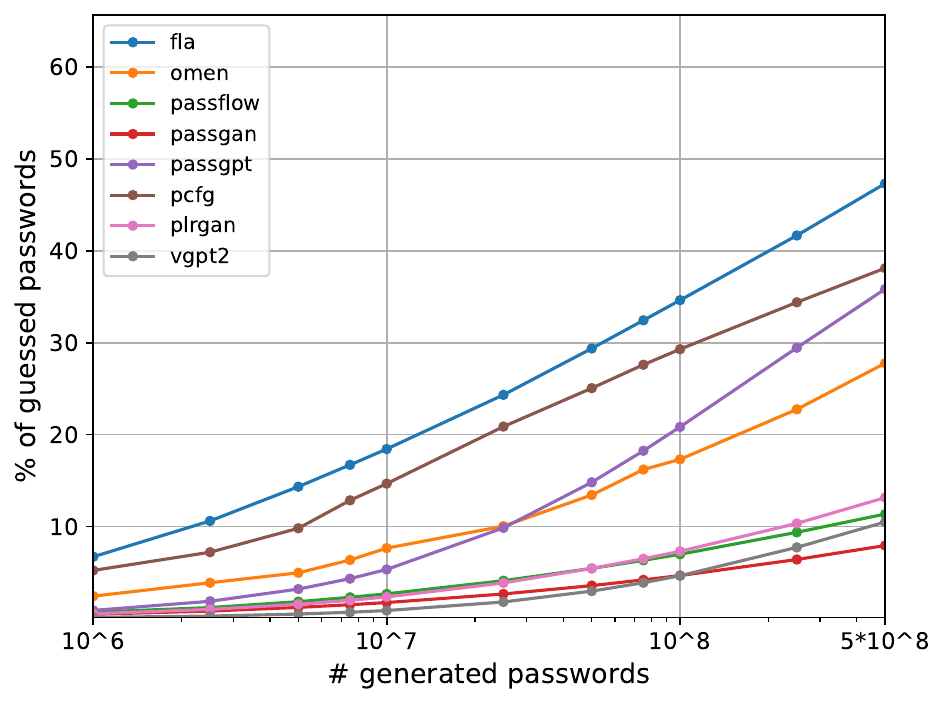}
    }
    \subfloat[Max. Password Length 12]{
      \includegraphics[width=0.24\textwidth]{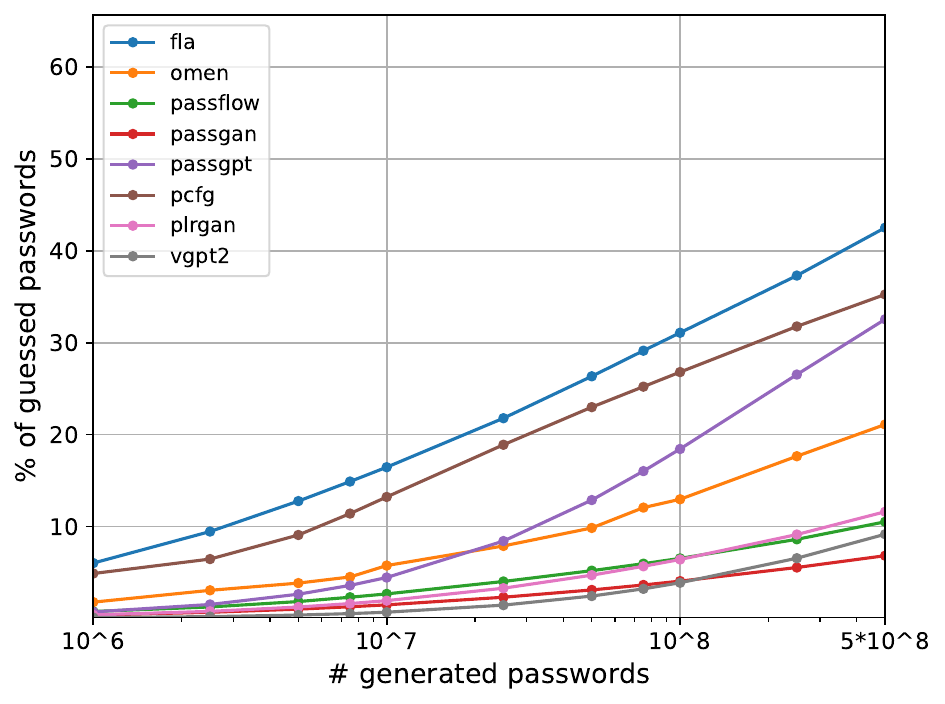}
    }       
    \caption{Impact of maximum password length on guessing performance over varying generation quantities.}
    \label{fig:rq1}
\end{figure*}

\subsubsection{\textbf{Frequency Distribution}} \label{sec:zipf}
Figure~\ref{fig:zipf-distributions} illustrates the frequency distribution of passwords for RockYou and 000webhost. To enhance readability and reduce the long-tail effect caused by less frequent passwords, we focus on passwords appearing at least three times in the dataset. A clear trend clearly emerges: a small subset of passwords are highly common, while the majority are rare. This behavior is characteristic of a Zipf-like distribution, where the frequency of a sample is inversely proportional to its rank. This observation aligns with the findings of Wang et al.~\cite{zipf_password_1}, who first demonstrated that Zipf’s law accurately models the distribution of human-chosen passwords.
\begin{table*}[t!]
    \centering
    \caption{Marginal gain in successful matches as the number of generated passwords increases. Each column represents a generation interval (e.g., 1M-2.5M indicates the marginal gain from 1M to 2.5M guesses). \textit{Total} reports the marginal gain from X to Y, expressed as a percentage of the total number of test passwords. \textit{Relative} indicates the marginal gain relative to the number of matches obtained at X. Results are averaged across all password lengths.}.
    \label{tab:marginal_gain}
    \renewcommand{\arraystretch}{1.15}

    \begin{adjustbox}{width=1\textwidth, center}
    \begin{tabular}{l *{20}{c}}
        \toprule
        \multirow{2}{*}{\textbf{Model}} & \multicolumn{2}{c}{\textbf{1M-2.5M}} & \multicolumn{2}{c}{\textbf{2.5M-5M}} & \multicolumn{2}{c}{\textbf{5M-7.5M}} & \multicolumn{2}{c}{\textbf{7.5M-10M}} & \multicolumn{2}{c}{\textbf{10M-25M}} & \multicolumn{2}{c}{\textbf{25M-50M}} & \multicolumn{2}{c}{\textbf{50M-75M}} & \multicolumn{2}{c}{\textbf{75M-100M}} & \multicolumn{2}{c}{\textbf{100M-250M}} & \multicolumn{2}{c}{\textbf{250M-500M}} \\
        \cmidrule(lr){2-3} \cmidrule(lr){4-5} \cmidrule(lr){6-7} \cmidrule(lr){8-9} \cmidrule(lr){10-11} \cmidrule(lr){12-13} \cmidrule(lr){14-15} \cmidrule(lr){16-17} \cmidrule(lr){18-19} \cmidrule(lr){20-21}
        & {\textbf{Total}} & {\textbf{Relative}} & {\textbf{Total}} & {\textbf{Relative}} & {\textbf{Total}} & {\textbf{Relative}} & {\textbf{Total}} & {\textbf{Relative}} & {\textbf{Total}} & {\textbf{Relative}} & {\textbf{Total}} & {\textbf{Relative}} & {\textbf{Total}} & {\textbf{Relative}} & {\textbf{Total}} & {\textbf{Relative}} & {\textbf{Total}} & {\textbf{Relative}} & {\textbf{Total}} & {\textbf{Relative}} \\
        \midrule
                FLA      & +4.33 & +58.54 & +4.17 & +35.55 & +2.66 & +16.71 & +1.93 & +10.40 & +6.57 & +32.14 & +5.55 & +20.62 & +3.33 & +10.28 & +2.34 & +6.56 & +7.44 & +19.62 & +5.84 & +13.00 \\
        OMEN      & +1.50 & +66.31 & +1.20 & +30.96 & +1.20 & +23.46 & +1.56 & +25.29 & +2.46 & +32.46 & +4.02 & +37.08 & +2.94 & +20.95 & +1.17 & +6.94 & +7.25 & +38.16 & +6.05 & +22.75 \\
        PassFlow & +0.62 & +69.67 & +0.73 & +48.88 & +0.56 & +25.15 & +0.44 & +15.72 & +1.72 & +52.72 & +1.58 & +31.53 & +1.05 & +15.85 & +0.80 & +10.38 & +2.84 & +33.66 & +2.48 & +21.97 \\
        PassGAN  & +0.49 & +83.75 & +0.55 & +50.94 & +0.40 & +24.84 & +0.32 & +15.95 & +1.23 & +53.74 & +1.14 & +32.85 & +0.75 & +16.47 & +0.56 & +10.69 & +1.96 & +34.15 & +1.63 & +21.38 \\
        PassGPT  & +1.20 & +114.72 & +1.60 & +71.91 & +1.34 & +35.32 & +1.18 & +22.97 & +5.21 & +83.58 & +5.45 & +48.49 & +3.66 & +22.15 & +2.73 & +13.56 & +8.89 & +39.13 & +6.53 & +20.63 \\
        PCFG      & +2.08 & +37.87 & +3.19 & +42.00 & +2.85 & +26.96 & +1.88 & +14.08 & +6.27 & +41.03 & +4.30 & +19.99 & +2.40 & +9.33 & +1.68 & +5.95 & +5.09 & +17.06 & +3.58 & +10.25 \\
        PLR-GAN  & +0.62 & +97.50 & +0.74 & +59.45 & +0.57 & +28.94 & +0.48 & +18.91 & +1.94 & +65.41 & +1.96 & +40.34 & +1.34 & +19.54 & +1.03 & +12.55 & +3.67 & +40.28 & +3.21 & +25.59 \\
        VGPT2    & +0.18 & +137.88 & +0.28 & +90.08 & +0.25 & +43.08 & +0.24 & +28.34 & +1.18 & +110.51 & +1.45 & +65.38 & +1.11 & +30.52 & +0.92 & +19.44 & +3.65 & +65.02 & +3.43 & +37.03 \\
        \bottomrule
    \end{tabular}
    \end{adjustbox}
\end{table*}

\section{\textbf{Experiments}}\label{sec:experiments}
We provide a thorough characterization of each selected model (Sections~\ref{sec:experiments-rq1}–\ref{sec:experiments-rq7}), addressing the research questions from Section~\ref{sec:framework-scenarios}, and laying the groundwork for the benchmark in Section~\ref{sec:experiments-benchmark}.

\subsection{\textbf{RQ1 - How Do Different Settings Influence Models Performance?}}\label{sec:experiments-rq1}
We examine the impact of two key parameters---maximum password length and the number of generated passwords---on model performance. Specifically, we compute the weighted average percentage of guessed passwords across all datasets, considering maximum lengths of 8, 10, and 12 characters. For each length, we evaluate 11 generation quantities, ranging from $10^6$ to $5 \times 10^8$ passwords. Results are shown in Figure~\ref{fig:rq1}. As expected, performance declines with increasing password length. However, this decline exhibits diminishing returns: while the drop from 8 to 10 characters is substantial, \rev{averaging 10.71 percentage points}, the decrease from 10 to 12 characters is markedly smaller, \rev{at just 2.80 points}, indicating a plateau as length increases. 
Among all models, FLA consistently outperforms the others, successfully guessing a substantial portion of passwords with relatively few guesses. Notably, FLA requires fewer than $10^7$ generated passwords to outperform all other models except PassGPT, \rev{PCFG, and OMEN}. PassGPT, while initially comparable to PLR-GAN, PassFlow, and PassGAN, exhibits a much steeper growth curve, \rev{growing even faster than} FLA after $5 \times 10^7$ \rev{passwords}. When considering few guesses, PLR-GAN, PassFlow, and PassGAN exhibit nearly identical performance, while VGPT2 initially underperforms. As the number of guesses increases, differences between the models become more pronounced: PLR-GAN begins to outperform PassFlow, and VGPT2, despite a slow start, improves faster than both. In contrast, PassGAN gains the least from additional guesses, with a relatively slow improvement rate. \rev{OMEN and PCFG, perform surprisingly well. OMEN initially follows a curve similar to PassGPT but later falls behind, especially for longer passwords. PCFG’s curve initially follows FLA’s, but its performance gains diminish as more guesses are generated. At $5 \times 10^8$ guesses, PCFG performs worse than PassGPT and OMEN for length 8, and slightly better than PassGPT for lengths 10 and 12.}

While it is evident that generating more passwords leads to an increased number of matches, we delve deeper into this trend by analyzing the \textit{marginal gain}---defined as the percentage increase in guessed passwords between two generation intervals. We analyze both \textit{total gains} (relative to the overall number of correct matches) and \textit{relative gains} (relative to the previous number of matches). Detailed results are reported in Table~\ref{tab:marginal_gain}.
Overall, all models exhibit clear diminishing returns, with sub-linear growth in successful guesses. Match rates grow rapidly in the early stages, then taper off as the space of common passwords is exhausted. Since this pattern is consistent across all lengths, and considering that over $92\%$ of passwords are 12 characters or fewer (see Table~\ref{tab:len_distribution}), we limit our subsequent analysis to this length.

\begin{figure*}[t!]
    \centering
    \subfloat[Rockyou]{
      \includegraphics[width=0.24\textwidth]{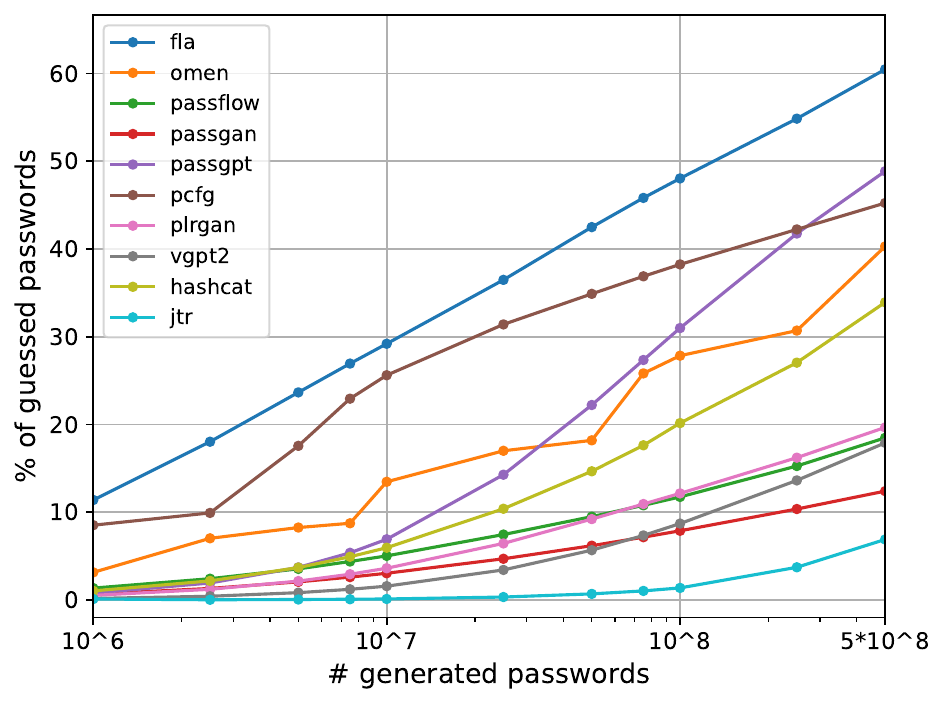}
    }
    \subfloat[Linkedin]{
      \includegraphics[width=0.24\textwidth]{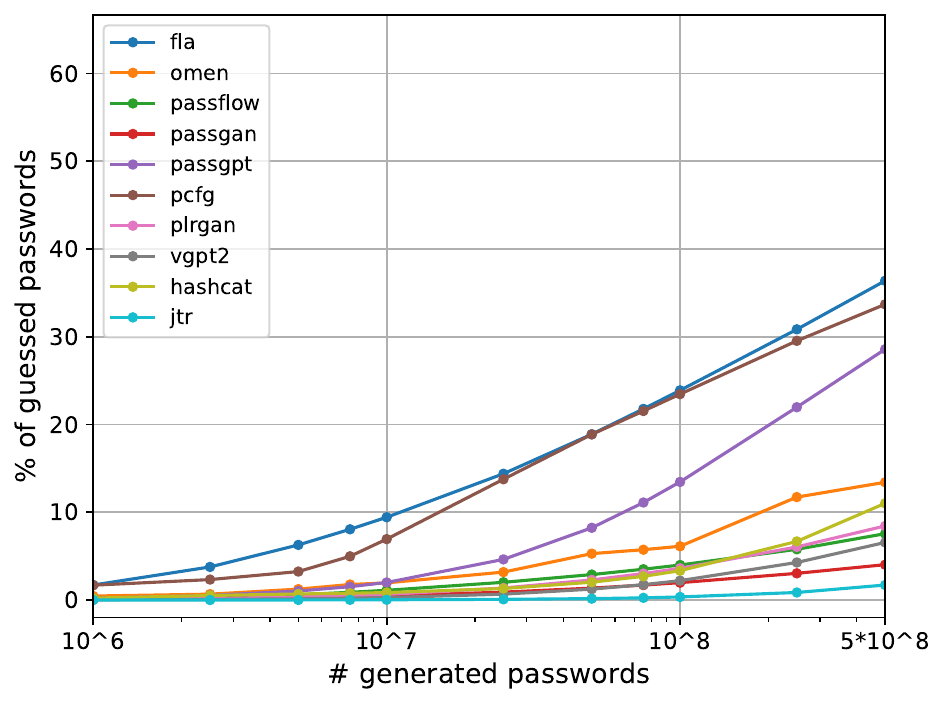}
    }    
    \subfloat[000webhost]{
      \includegraphics[width=0.24\textwidth]{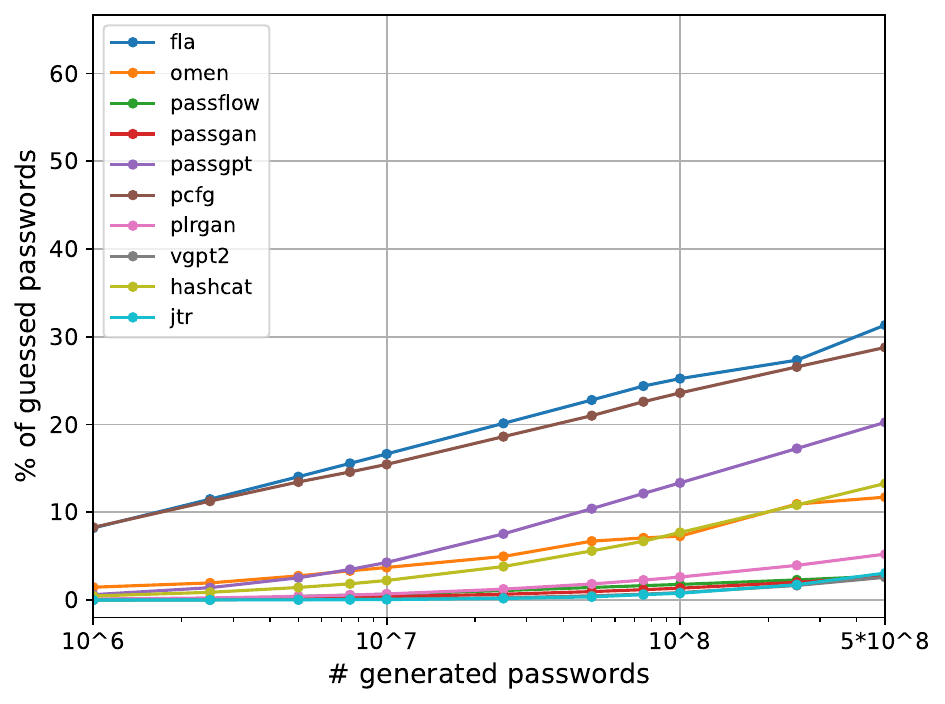}
    }    
    \subfloat[Ashley Madison]{
      \includegraphics[width=0.24\textwidth]{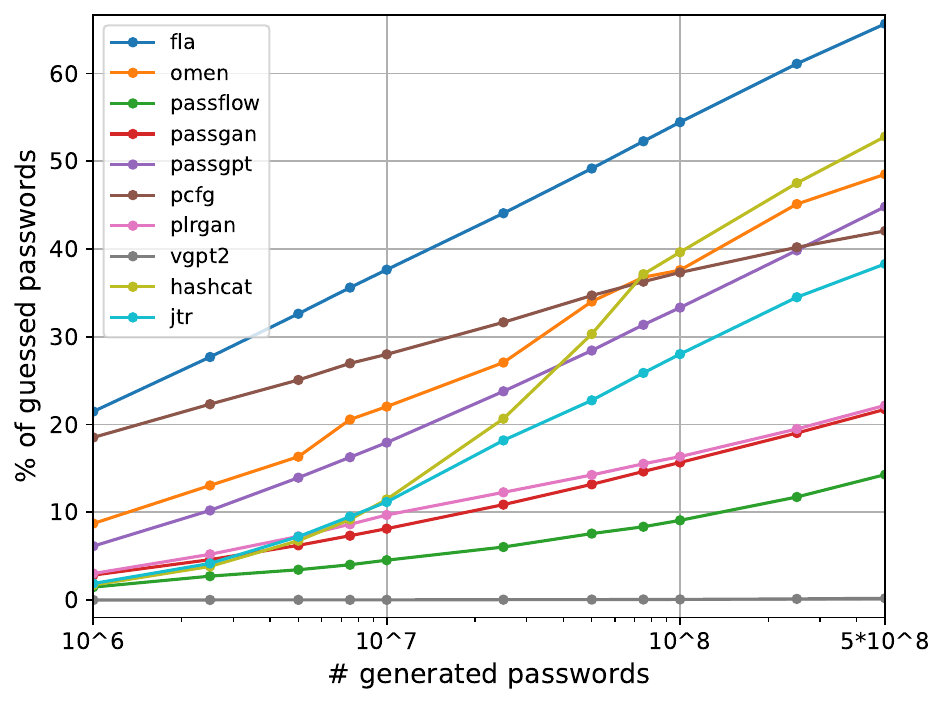}
    }   
    \\
    \subfloat[Gmail]{
      \includegraphics[width=0.24\textwidth]{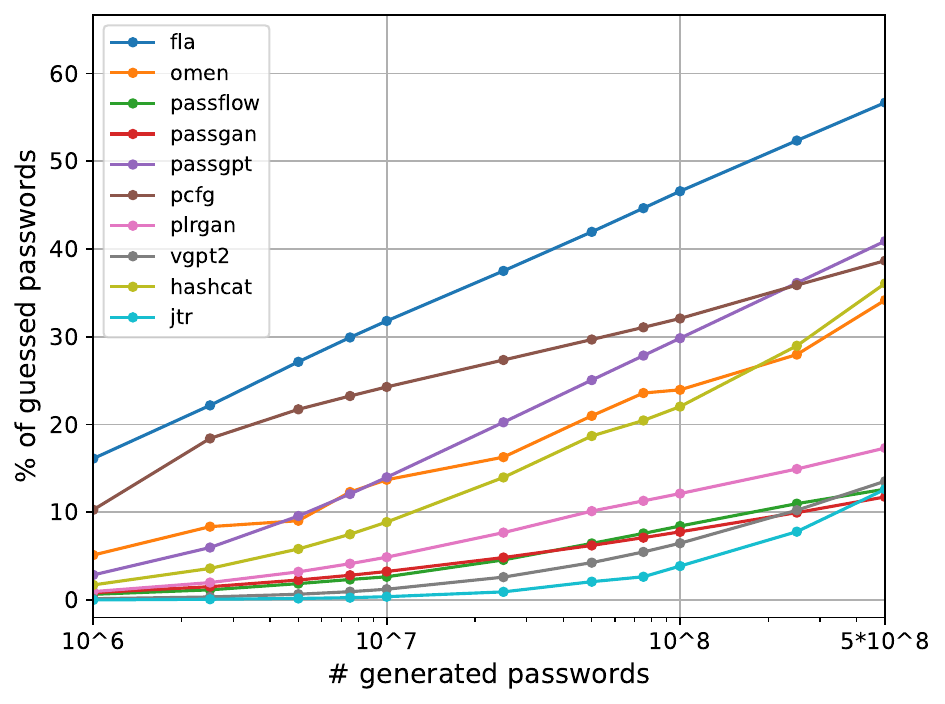}
    } 
    \subfloat[Mailru]{
      \includegraphics[width=0.24\textwidth]{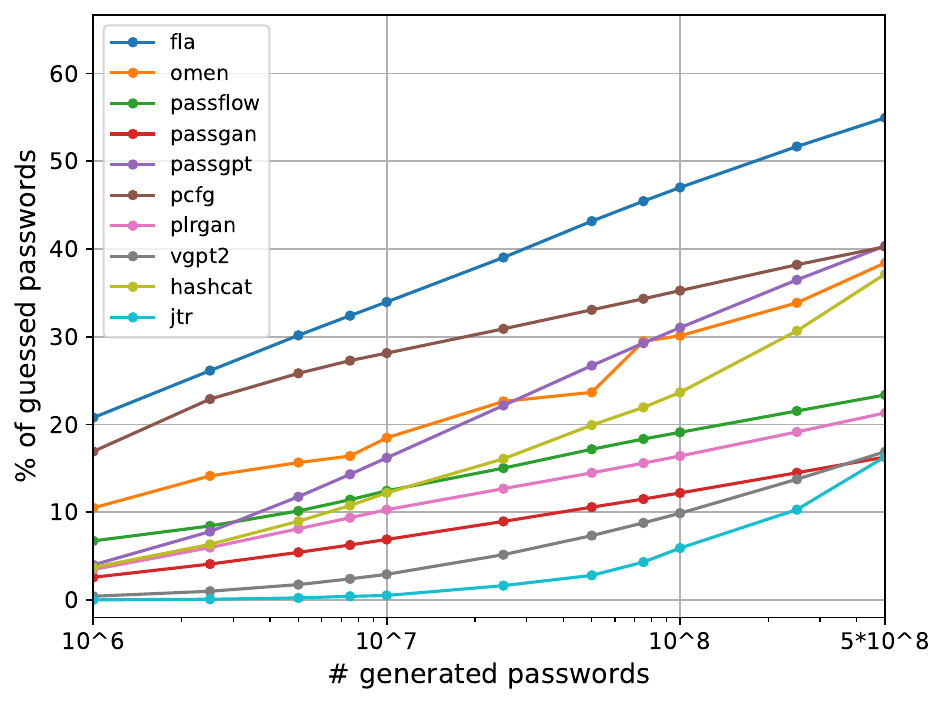}
    }    
    \subfloat[Taobao]{
      \includegraphics[width=0.24\textwidth]{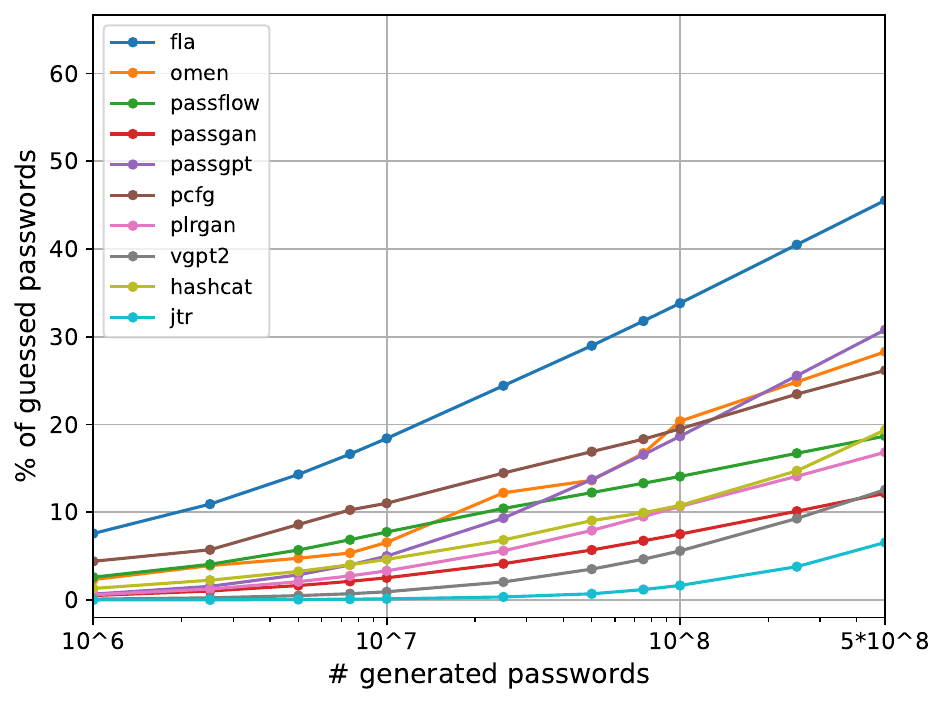}
    }   
    \subfloat[Libero]{
      \includegraphics[width=0.24\textwidth]{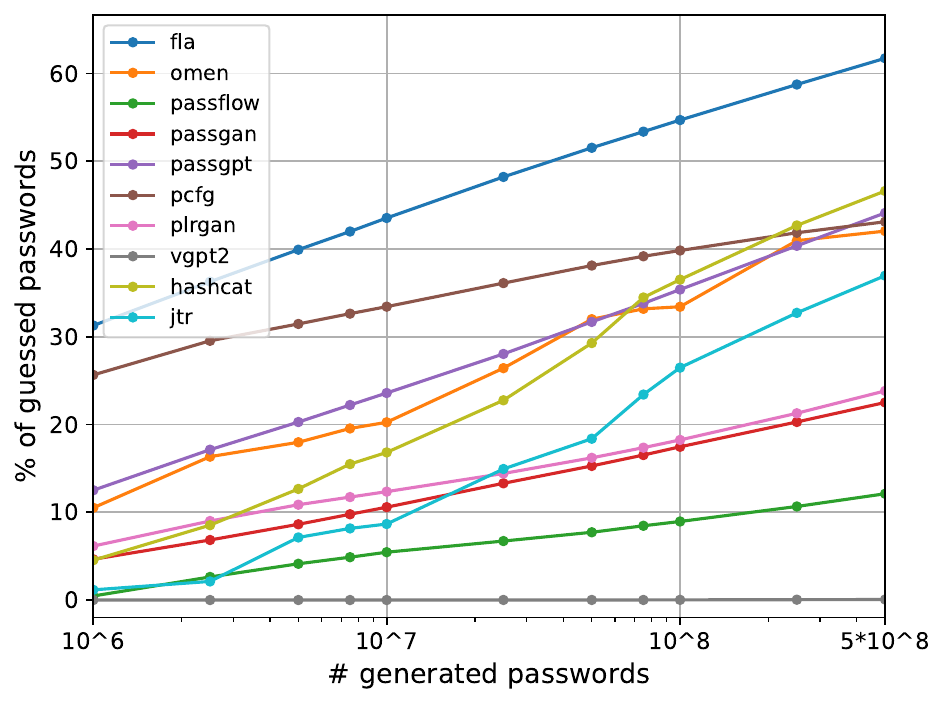}
    }   
    \caption{Comparison between traditional methods and generative models across 8 datasets.}
    \label{fig:rq2}
\end{figure*}

\begin{table*}[t!]
    \centering
    \caption{Impact of training dataset size on performance. Values expressed as percentage of guessed test set passwords.}
    \label{tab:rq3}
    \renewcommand{\arraystretch}{1} 
    \begin{adjustbox}{width=1\textwidth, center}
        \begin{tabular}{l *{32}{c}}
            \toprule
            \multirow{2}{*}{\textbf{Size}} 
              & \multicolumn{8}{c}{\textbf{Mailru}} 
              & \multicolumn{8}{c}{\textbf{Taobao}} 
              & \multicolumn{8}{c}{\textbf{RockYou}} 
              & \multicolumn{8}{c}{\textbf{LinkedIn}} \\
            \cmidrule(lr){2-9} \cmidrule(lr){10-17} \cmidrule(lr){18-25} \cmidrule(lr){26-33}
            & {FLA} & {OMEN} & {PFLW} & {PGAN} & {PGPT} & {PCFG} & {PLR} & {VGPT} 
            & {FLA} & {OMEN} & {PFLW} & {PGAN} & {PGPT} & {PCFG} & {PLR} & {VGPT} 
            & {FLA} & {OMEN} & {PFLW} & {PGAN} & {PGPT} & {PCFG} & {PLR} & {VGPT} 
            & {FLA} & {OMEN} & {PFLW} & {PGAN} & {PGPT} & {PCFG} & {PLR} & {VGPT} \\
            \midrule
        1e6 & 53.14 & 37.78 & 23.27 & 12.73 & 32.93 & 36.24 & 21.26 & 6.25
            & 42.94 & 27.90 & 17.58 & 13.15 & 27.51 & 19.10 & 16.45 & 4.24
            & 56.13 & 38.70 & 18.21 & 9.60 & 30.98  & 32.80 & 19.29 & 5.97
            & 18.48 & 12.68 & 7.06 & 4.42 & 12.49   & 22.89 & 7.83 & 1.48 \\
        2e6 & 54.95 & 38.42 & 23.24 & 16.16 & 40.35 & 40.26 & 21.30 & 16.90
            & 44.12 & 28.18 & 18.65 & 10.16 & 29.11 & 22.05 & 16.29 & 11.28
            & 57.36 & 39.65 & 18.35 & 11.31 & 41.72 & 36.67 & 19.21 & 13.75
            & 24.61 & 13.07 & 7.12 & 4.38 & 17.97 & 25.33 & 8.08 & 4.44 \\
        3e6 & {--} & {--} & {--} & {--} & {--} & {--} & {--} & {--} 
            & 44.79 & 28.19 & 17.21 & 10.80 & 29.81 & 23.75 & 16.05 & 11.37
            & 60.45 & 39.93 & 19.32 & 12.78 & 44.01 & 38.71 & 17.69 & 15.25
            & 29.30 & 13.22 & 6.88 & 3.79 & 19.75 & 26.45 & 7.92 & 4.55 \\
        5e6 & {--} & {--} & {--} & {--} & {--} & {--} & {--} & {--} 
            & 45.53 & 28.29 & 18.49 & 12.11 & 30.80 & 26.17 & 16.82 & 12.56
            & 60.08 & 40.11 & 19.35 & 12.22 & 46.40 & 41.27 & 18.06 & 16.49
            & 37.89 & 13.32 & 6.84 & 2.52 & 22.00 & 27.80 & 8.38 & 5.38 \\
        1e7 & {--} & {--} & {--} & {--} & {--} & {--} & {--} & {--} 
            & {--} & {--} & {--} & {--} & {--} & {--} & {--} & {--} 
            & 60.47 & 40.29 & 18.46 & 12.36 & 48.85 & 45.24 & 19.65 & 17.90
            & 41.05 & 13.37 & 7.00 & 3.33 & 25.27 & 30.37 & 6.53 & 5.97 \\
        2e7 & {--} & {--} & {--} & {--} & {--} & {--} & {--} & {--} 
            & {--} & {--} & {--} & {--} & {--} & {--} & {--} & {--} 
            & {--} & {--} & {--} & {--} & {--} & {--} & {--} & {--} 
            & 35.10 & 13.42 & 7.13 & 4.87 & 26.91 & 31.94 & 7.12 & 6.33 \\
        4e7 & {--} & {--} & {--} & {--} & {--} & {--} & {--} & {--} 
            & {--} & {--} & {--} & {--} & {--} & {--} & {--} & {--} 
            & {--} & {--} & {--} & {--} & {--} & {--} & {--} & {--} 
            & 36.37 & 13.42 & 7.10 & 3.99 & 28.58 & 33.69 & 8.44 & 6.56 \\
            \bottomrule
        \end{tabular}
    \end{adjustbox}
\end{table*}

\subsection{\textbf{RQ2 - Are \rev{Deep} Generative Models Truly Better Than Traditional Tools?}}

We compare \rev{DL-based generative} models against two categories of traditional password-guessing techniques: (1) rule-based attacks, namely, Hashcat and John the Ripper, both operating in wordlist mode, using 'Unicorn Rules' and 'Wordlist' rulesets, respectively; and (2) \rev{ML-based generative} approaches, specifically PCFG and OMEN. Additional details on these tools are provided in Appendix~\ref{sec:appendix-models}. 

As shown in Figure~\ref{fig:rq2}, rule-based tools demonstrate strong performance on smaller datasets, such as Ashley Madison and Libero. In these cases, Hashcat outperforms all models except FLA. Both OMEN and PCFG achieve results comparable to PassGPT, successfully guessing a significant portion of the test passwords, while JtR follows closely behind.

However, as dataset size increases, the advantage shifts toward \rev{learning-based} models, with the performance gap over rule-based tools widening considerably. On average, Hashcat underperforms compared to the top four learning-based approaches, while JtR consistently ranks as the least effective. We further investigated the rationale behind this declining performance, hypothesizing it stems from their underlying generation strategies. Unicorn Rules are ordered by efficacy, and we selected the first $n$ passwords generated by Hashcat accordingly. However, Hashcat applies a rule-first strategy: it processes all rules for the current password before moving to the next. As the dataset size increases, a smaller portion of the test set passwords are transformed, concentrating Hashcat's guesses in a small subset of the overall password space. JtR follows a password-first strategy, but WordList's rules are unsorted, so we randomly sampled $n$ passwords from the generated set. In larger datasets, the number of generated candidates increases substantially, diluting the proportion of successful guesses and making it less likely to sample correct guesses. 

On the two most challenging datasets, LinkedIn and 000webhost, where all models exhibit the lowest guess rates, PCFG closely matches FLA’s performance, trailing by only a few percentage points. PassGPT is the only other approach nearing their effectiveness, while Hashcat and OMEN show comparable but noticeably lower performance. Notably, except for these two datasets, PCFG successfully guesses a large number of passwords in the early stages, nearly matching FLA, but its guessing curve increases more slowly over time, likely due to early saturation. 

Overall, learning-based approaches outperform traditional tools, especially as scale and complexity grow. Among them, on average, FLA and PassGPT lead, while older ML techniques like PCFG still show strong performance.

To streamline subsequent analyses, we fix the number of generated passwords at $5 \times 10^8$, as it yields the best results.

\begin{table*}[t!]
    \centering
    \caption{Cross-community generalization ability. Values expressed as percentage of guessed test set passwords.}
    \label{tab:rq4.1}
    \renewcommand{\arraystretch}{1.15}

    \begin{adjustbox}{width=1\linewidth, center}
    
    \begin{tabular}{l *{24}{c}}
        \toprule
        \multirow{2}{*}{\textbf{Train / Test}} & \multicolumn{3}{c}{\textbf{FLA}} & \multicolumn{3}{c}{\textbf{OMEN}} & \multicolumn{3}{c}{\textbf{PassFlow}} & \multicolumn{3}{c}{\textbf{PassGAN}} & \multicolumn{3}{c}{\textbf{PassGPT}} & \multicolumn{3}{c}{\textbf{PCFG}} & \multicolumn{3}{c}{\textbf{PLR-GAN}} & \multicolumn{3}{c}{\textbf{VGPT2}} \\
        \cmidrule(lr){2-4} \cmidrule(lr){5-7} \cmidrule(lr){8-10} \cmidrule(lr){11-13} \cmidrule(lr){14-16} \cmidrule(lr){17-19} \cmidrule(lr){20-22} \cmidrule(lr){23-25}
         & {\textbf{000W.}} & {\textbf{Link.}} & {\textbf{Rock.}} & {\textbf{000W.}} & {\textbf{Link.}} & {\textbf{Rock.}} & {\textbf{000W.}} & {\textbf{Link.}} & {\textbf{Rock.}} & {\textbf{000W.}} & {\textbf{Link.}} & {\textbf{Rock.}} & {\textbf{000W.}} & {\textbf{Link.}} & {\textbf{Rock.}} & {\textbf{000W.}} & {\textbf{Link.}} & {\textbf{Rock.}} & {\textbf{000W.}} & {\textbf{Link.}} & {\textbf{Rock.}} & {\textbf{000W.}} & {\textbf{Link.}} & {\textbf{Rock.}} \\
        \midrule
        000Webhost & 31.33 & 22.81 & 26.72 & 11.72 & 7.55 & 9.63 & 2.74 & 6.58 & 11.44 & 2.66 & 2.27 & 4.10 & 20.24 & 10.41 & 13.77 & 28.78 & 20.30 & 23.93 & 5.22 & 3.70 & 5.56 & 2.59 & 1.87 & 3.15 \\
        LinkedIn & 19.01 & 36.37 & 45.09 & 7.53 & 13.42 & 17.90 & 1.93 & 7.10 & 6.61 & 1.80 & 4.03 & 6.51 & 16.90 & 28.58 & 36.21 & 23.99 & 33.69 & 40.30 & 3.65 & 8.44 & 8.77 & 2.95 & 6.56 & 11.48 \\
        RockYou & 17.31 & 31.53 & 60.47 & 8.10 & 17.60 & 40.29 & 3.55 & 8.16 & 18.46 & 1.59 & 4.72 & 12.41 & 13.15 & 22.08 & 48.85 & 20.75 & 27.17 & 45.24 & 4.65 & 8.77 & 19.67 & 2.84 & 6.93 & 17.90 \\
        \bottomrule
    \end{tabular}
    \end{adjustbox}
\end{table*}

\begin{table*}[t!]
    \centering
    \caption{Cross-culture generalization ability. Values expressed as percentage of guessed test set passwords.}
    \label{tab:rq4.2}
    \renewcommand{\arraystretch}{1.15}
    \begin{adjustbox}{width=1\linewidth, center}
    \begin{tabular}{l *{24}{c}}
        \toprule
        \multirow{2}{*}{\textbf{Train / Test}} & \multicolumn{3}{c}{\textbf{FLA}} & \multicolumn{3}{c}{\textbf{OMEN}} & \multicolumn{3}{c}{\textbf{PassFlow}} & \multicolumn{3}{c}{\textbf{PassGAN}} & \multicolumn{3}{c}{\textbf{PassGPT}} & \multicolumn{3}{c}{\textbf{PCFG}} & \multicolumn{3}{c}{\textbf{PLR-GAN}} & \multicolumn{3}{c}{\textbf{VGPT2}} \\
        \cmidrule(lr){2-4} \cmidrule(lr){5-7} \cmidrule(lr){8-10} \cmidrule(lr){11-13} \cmidrule(lr){14-16} \cmidrule(lr){17-19} \cmidrule(lr){20-22} \cmidrule(lr){23-25}
         & {\textbf{Mail.}} & {\textbf{Rock.}} & {\textbf{Taob.}} & {\textbf{Mail.}} & {\textbf{Rock.}} & {\textbf{Taob.}} & {\textbf{Mail.}} & {\textbf{Rock.}} & {\textbf{Taob.}} & {\textbf{Mail.}} & {\textbf{Rock.}} & {\textbf{Taob.}} & {\textbf{Mail.}} & {\textbf{Rock.}} & {\textbf{Taob.}} & {\textbf{Mail.}} & {\textbf{Rock.}} & {\textbf{Taob.}} & {\textbf{Mail.}} & {\textbf{Rock.}} & {\textbf{Taob.}} & {\textbf{Mail.}} & {\textbf{Rock.}} & {\textbf{Taob.}} \\
        \midrule
        Mailru & 54.95 & 26.98 & 16.36 & 38.42 & 15.89 & 12.29 & 23.37 & 16.35 & 13.25 & 16.29 & 7.39 & 4.72 & 40.35 & 15.82 & 10.74 & 40.26 & 14.57 & 6.37 & 21.32 & 10.78 & 7.11 & 16.90 & 6.89 & 4.11 \\
        RockYou & 30.10 & 60.47 & 18.71 & 19.43 & 40.29 & 16.28 & 14.86 & 18.48 & 9.83 & 8.43 & 12.41 & 5.57 & 22.30 & 48.85 & 13.65 & 23.25 & 45.24 & 9.54 & 13.22 & 19.67 & 9.10 & 11.40 & 17.90 & 6.52 \\
        Taobao & 20.94 & 23.77 & 45.53 & 11.11 & 10.05 & 28.29 & 20.20 & 19.55 & 18.68 & 7.39 & 6.41 & 12.16 & 13.61 & 13.26 & 30.80 & 10.64 & 11.72 & 26.17 & 10.11 & 9.37 & 16.84  & 8.16 & 7.74 & 12.56 \\
        \bottomrule
    \end{tabular}
    \end{adjustbox}
\end{table*}

\subsection{\textbf{RQ3 - How Sensitive are Models to Training Dataset Size?}}
We assess the ability of \rev{the} models to capture the full password distribution when trained on different data subset sizes. Models were trained on up to seven different subset sizes across four datasets, with a minimum initial size of 1M passwords. Results in Table~\ref{tab:rq3} illustrate the models' varying performance. \rev{PCFG and both} transformer-based models (PassGPT and VGPT2) consistently improve as the training dataset size increases, with VGPT2, in particular, struggling on smaller subsets. FLA also generally benefits from larger training subsets, although the rate of improvement tapers off between $3e6$ and $1e7$ passwords. We also observe an anomalous behavior on LinkedIn, where performance drops significantly after $1e7$---or $25\%$ of the dataset size. PLR-GAN excels on small training subsets, but its performance declines around $30\%$ of the dataset size before improving again, ultimately achieving its best performance on the full dataset. \rev{OMEN and} PassFlow \rev{remain} the most consistent \rev{models} across subset sizes, with minimal performance variation. In contrast, PassGAN exhibits highly variable results, with performance peaks at different dataset sizes, making its behavior less predictable. 

Overall, transformer models show the most significant improvement with increasing training data, whereas other architectures tend to exhibit flat or minimal gains, challenging the conventional assumption that more training data universally leads to better performance.

\subsection{\textbf{RQ4 - Can Models Generalize To Different Communities and/or Cultures?}}
We investigate the generalization capabilities of generative models by evaluating them in two cross-dataset scenarios: (1) cross-community and (2) cross-culture. In the cross-community setting, we use three datasets---RockYou, 000webhost, and LinkedIn---that share the same language but represent distinct user communities. For the cross-cultural setting, we used RockYou, Mailru, and Taobao, which differ in language, cultural background, and community.

\subsubsection{\textbf{Cross-community}}
As shown in Table~\ref{tab:rq4.1}, Cross-community generalization is strongly model- and dataset-dependent. \rev{OMEN}, PassGAN, PLR-GAN, PassFlow, and VGPT2 achieve their highest performance on LinkedIn when trained on RockYou, suggesting that the distribution learned from RockYou generalizes well despite community differences. However, stronger models such as PassGPT, \rev{PCFG}, and FLA experience notable performance drops in the same scenario. This contrast leads us to hypothesize that the apparent generalization success of weaker models stems from their limited capacity to capture fine-grained details of the training distribution---resulting in broader, but less accurate, generalization. We also note how all approaches struggle to generalize when training on 000webhost, highlighting that performance significantly degrades when the source and target distributions diverge too strongly (see Table~\ref{tab:pattern_distribution}).

\subsubsection{\textbf{Cross-culture}}
In the cross-culture scenario, model performance follows a more expected pattern, as illustrated in Table~\ref{tab:rq4.2}: models trained on one cultural or linguistic context tend to perform significantly worse when evaluated on datasets from a different one. This highlights the challenges of generalizing across culturally distinct password distributions and underscores the importance of training data alignment with the target population. However, some interesting behaviors still emerge. Despite the challenging setting, models still manage to guess a non-negligible portion of the target passwords. Notably, \rev{generative} models that typically underperform in other settings—such as PassFlow—can surpass stronger models like PassGPT, \rev{OMEN, and PCFG,} when there is a significant mismatch between source and target distributions. This suggests that certain models may possess greater flexibility or robustness in the face of distributional shifts, even if they are less effective under ideal, in-distribution conditions.

\begin{figure}[t!]
    \centering
    \includegraphics[width=0.67\columnwidth]{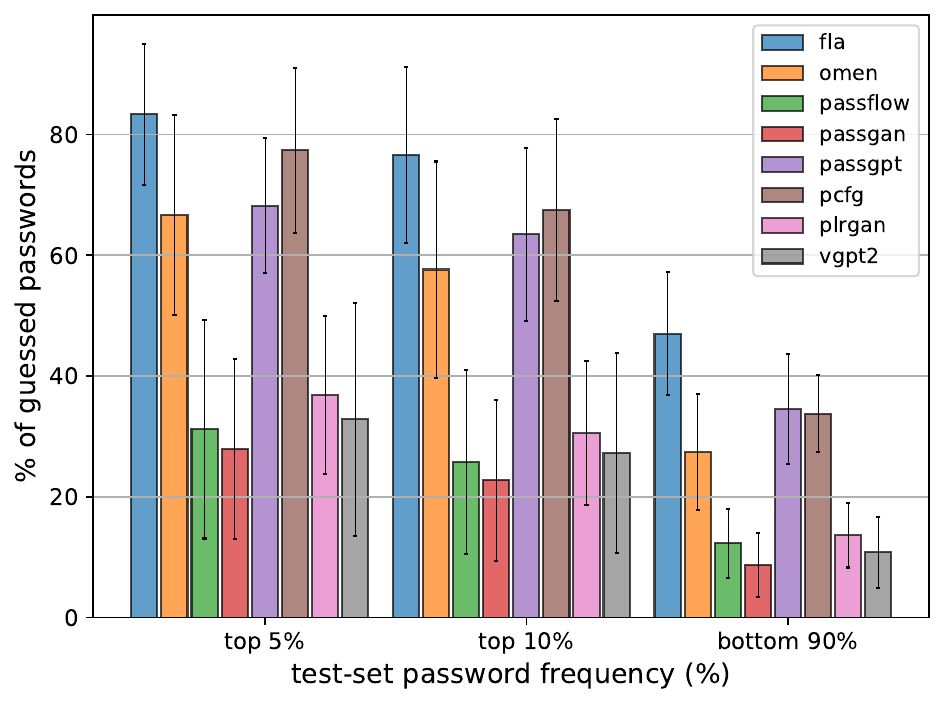}
    \caption{Guessing performance by password frequency.}     
    \label{fig:rq5}
\end{figure}

\subsection{\textbf{RQ5 - Are Models Limited to Guessing Only Simple and Common Passwords?}}
We evaluate the models’ ability to guess both common/rare and simple/complex passwords through three complementary analyses.

\subsubsection{\textbf{Analysis by Password Frequency}}
\begin{figure}[t!]
    \centering
    \subfloat[RockYou]{
      \includegraphics[width=0.48\columnwidth]{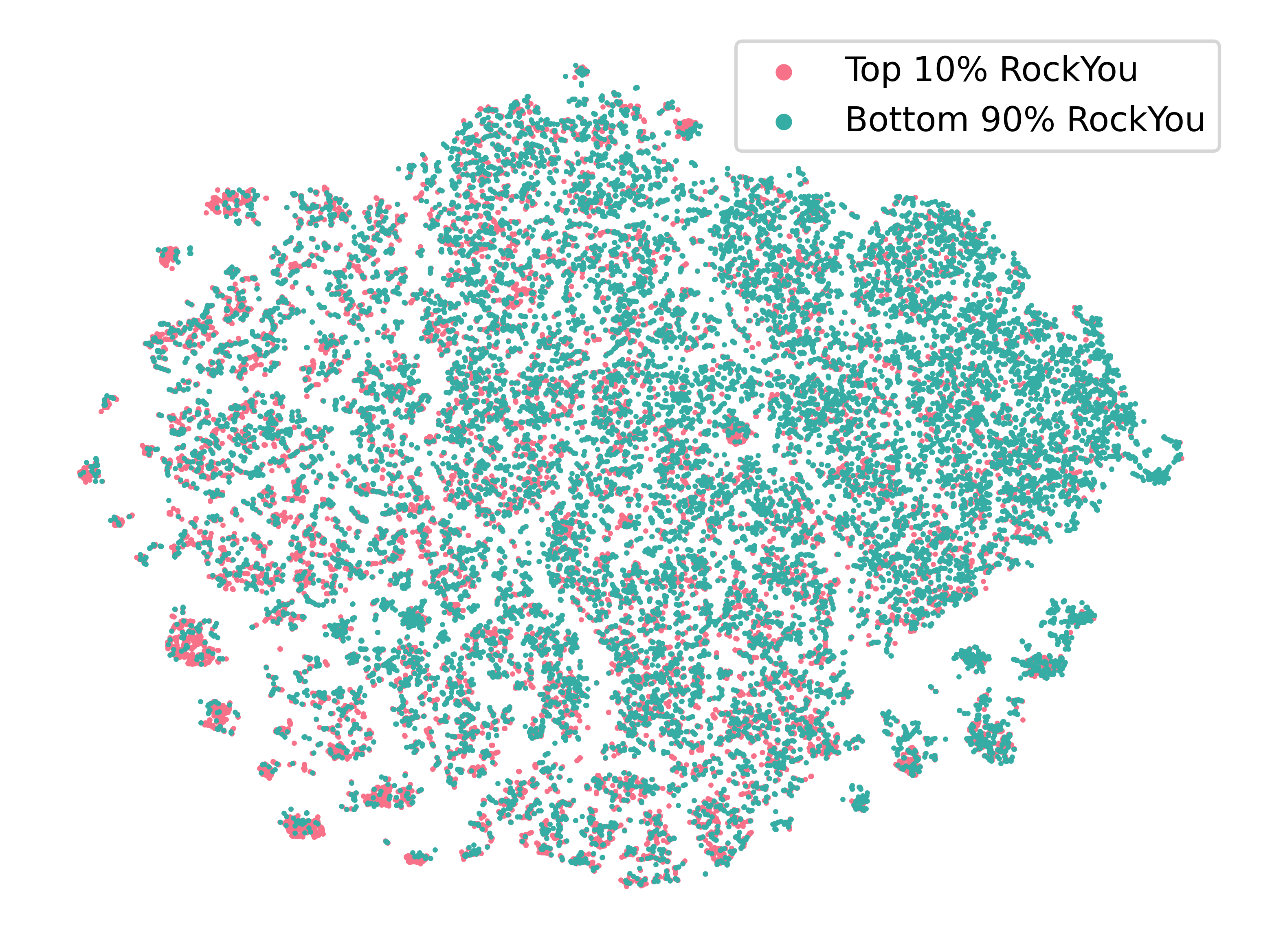}
    }
    \subfloat[Gmail]{
      \includegraphics[width=0.48\columnwidth]{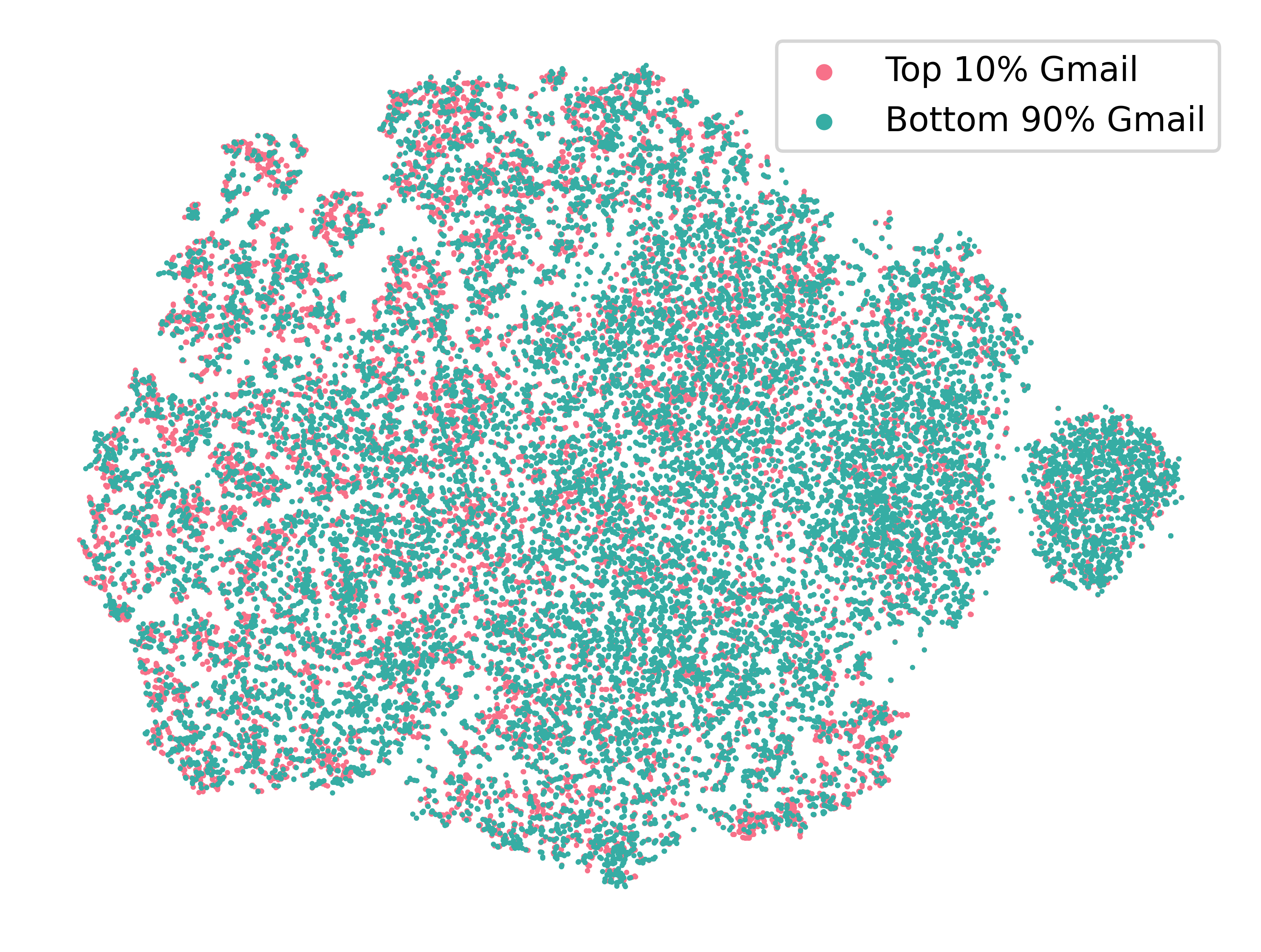}
    }
    \caption{t-SNE plot showing the projection of Top 10\% and Bottom 90\% passwords in PassFlow's latent space.}
    \label{fig:rq5-embed}
\end{figure}

From each test dataset, we created three subsets based on password frequency: top $5\%$, top $10\%$, and bottom $90\%$. Since nearly $90\%$ of test passwords across all datasets are unique (see Figure~\ref{fig:zipf-distributions}), these subsets allow us to evaluate models under three scenarios: (1) very common, (2) common, and (3) rare passwords. We excluded LinkedIn and Ashley Madison, as they consist \rev{nearly} entirely of unique passwords \rev{to avoid biased findings}. Results, shown in Figure~\ref{fig:rq5}, are reported as weighted averages. As expected, all models perform better on common passwords. While the drop from the top $5\%$ to the top $10\%$ is small, performance declines sharply when focusing on the bottom $90\%$. Still, models are surprisingly capable of guessing rare passwords, with PassGPT \rev{and PCFG} guessing close to $40\%$, and FLA close to $50\%$. 
In Figure~\ref{fig:rq5-embed}, we leverage PassFlow’s encoder to visualize password embeddings and observe that the bottom $90\%$ of passwords tend to form clusters around the top $10\%$. Since PassFlow's latent space is smooth~\cite{passflow}, this behavior suggests that many rare passwords are slight variations of frequently used ones, which may explain why models are still able to guess a non-negligible portion of them despite their low frequency.

\begin{table*}[t!]
    \centering
    \caption{Guessing performance by password pattern. Values expressed as percentage of guessed passwords for each pattern.}
    \label{tab:rq5-matches_by_pattern}
    \renewcommand{\arraystretch}{1.15}

    \begin{adjustbox}{width=1\textwidth, center}
    \begin{tabular}{l *{19}{c}}
        \toprule
        \textbf{Model} & {\textbf{r1}} & {\textbf{r2}} & {\textbf{r3}} & {\textbf{r4}} & {\textbf{r5}} & {\textbf{r6}} & {\textbf{r7}} & {\textbf{r8}} & {\textbf{r9}} & {\textbf{r10}} & {\textbf{r11}} & {\textbf{r12}} & {\textbf{r13}} & {\textbf{r14}} & {\textbf{r15}} & {\textbf{r16}} & {\textbf{r17}} & {\textbf{r18}} & {\textbf{r19}} \\
        \midrule
        FLA & 46.51 & 48.14 & 38.49 & 67.56 & 17.51 & 39.38 & 21.02 & 26.28 & 14.20 & 44.63 & 21.32 & 26.39 & 11.88 & 59.87 & 13.92 & 8.50  & 4.58  & 24.57 & 48.09 \\
        OMEN & 23.43 & 24.94 & 11.03 & 55.07 & 13.13 & 15.97 & 4.21 & 6.69 & 2.92 & 18.8 & 4.13 & 6.72 & 3.12 & 47.41 & 2.53 & 0.63 & 0.99 & 4.48 & 23.41 \\ 
        PassFlow & 10.38 & 11.35 & 1.41  & 27.92 & 5.34  & 4.82  & 0.70  & 0.30  & 0.05  & 5.01  & 0.41  & 2.29  & 0.21  & 24.02 & 1.79  & 0.45  & 0.16  & 0.39  & 9.54  \\
        PassGAN  & 6.94  & 7.65  & 0.16  & 28.31 & 0.30  & 3.55  & 0.10  & 0.23  & 0.08  & 4.12  & 0.10  & 0.53  & 0.12  & 24.17 & 0.10  & 0.01  & 0.07  & 0.06  & 7.53  \\
        PassGPT  & 38.47 & 39.66 & 33.74 & 46.55 & 25.82 & 29.81 & 20.04 & 19.90 & 11.19 & 33.43 & 17.88 & 29.62 & 10.39 & 41.68 & 18.19 & 12.72 & 5.04  & 21.63 & 36.16 \\
        PCFG & 29.79 & 28.69 & 45.59 & 13.69 & 19.81 & 42.94 & 22.16 & 16.41 & 12.11 & 48.87 & 21.5 & 45.14 & 12.31 & 13.46 & 30.14 & 8.78 & 5.35 & 22.19 & 43.64 \\ 
        PLR-GAN  & 12.60 & 13.74 & 1.35  & 39.95 & 0.67  & 7.31  & 0.77  & 0.74  & 0.27  & 8.39  & 0.64  & 2.14  & 0.47  & 34.17 & 0.35  & 0.03  & 0.15  & 1.02  & 12.49 \\
        VGPT2 & 12.09 & 13.11 & 2.38  & 29.39 & 2.59  & 5.37  & 1.22  & 0.90  & 0.33  & 6.10  & 0.95  & 2.15  & 0.53  & 25.19 & 0.61  & 0.04  & 0.09  & 1.07  & 9.74  \\        \bottomrule
    \end{tabular}
    \end{adjustbox}
\end{table*}

\subsubsection{\textbf{Analysis by Password Length}}
\begin{figure}[t!]
\includegraphics[width=\columnwidth]{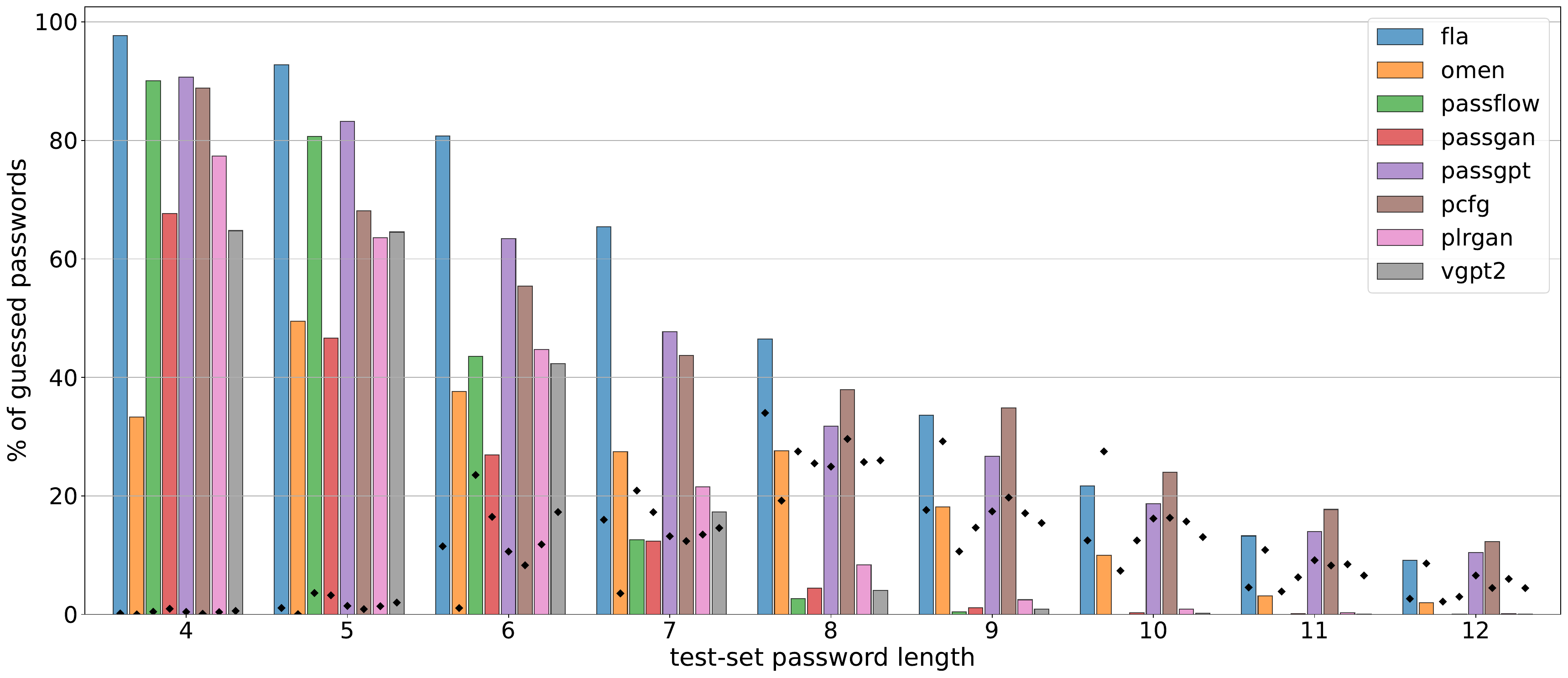}
    \caption{Guessing performance by password length. Dots show the percentage of generated passwords for each length.}
    \label{fig:rq5-match-by-length}
\end{figure}

We analyzed matches by password length, ranging from 4 to 12 characters. Results, computed using a weighted average across all datasets, are shown in Figure~\ref{fig:rq5-match-by-length}. \rev{All models, except OMEN}, perform well on short passwords, with FLA nearly achieving a $100\%$ match rate. However, performance declines as length increases. \rev{Beyond 8 characters, only PCFG, OMEN (up to 10 characters), FLA,} and PassGPT maintain a substantial guessing rate. \rev{From length 9 onward, PCFG emerges as the most performing approach, surpassing both FLA and PassGPT, while,} consistent with \textbf{RQ1}, PassGPT \rev{overtakes} FLA from length 11. PassFlow initially performs on par with PassGPT for shorter lengths but undergoes a sharp decline, becoming the weakest performer from length 8 onward.

\subsubsection{\textbf{Analysis by Password Patterns}}
We evaluated the models’ ability to guess passwords based on the structural patterns defined in Table~\ref{tab:regex}. The results, presented in Table~\ref{tab:rq5-matches_by_pattern} as a weighted average across all datasets, show that PassGPT, FLA, \rev{PCFG, and, in part, OMEN} consistently achieve the strongest performance, standing out as the only models capable of successfully guessing passwords across all defined patterns. In contrast, the remaining models show limited pattern coverage, particularly struggling with complex password structures that include special characters (r5, r7, r8) or uncommon character combinations (r11, r12, r13). GAN-based models fare especially poorly on passwords composed entirely of special characters (r5), significantly underperforming PassFlow and VGPT2. This suggests a clear limitation in capturing rare or unconventional character sets. In contrast, all models exhibit strong performance on digit-only passwords (r4), indicating that simple patterns are consistently learned regardless of the architecture.

\subsection{\textbf{RQ6 - To What Extent Do the Distributions Learned by Different Models Align? Can We Combine Models to Maximize Effectiveness?}}
We assess the extent to which approaches' learned distributions overlap and investigate the effectiveness of combining multiple models to enhance guessing performance.

\subsubsection{\textbf{RQ6.1 - To What Extent Do the Distributions Learned by Different Models Align?}}
\begin{figure}[t]
    \centering
    \subfloat[Jaccard Index]{\includegraphics[width=0.465\columnwidth]{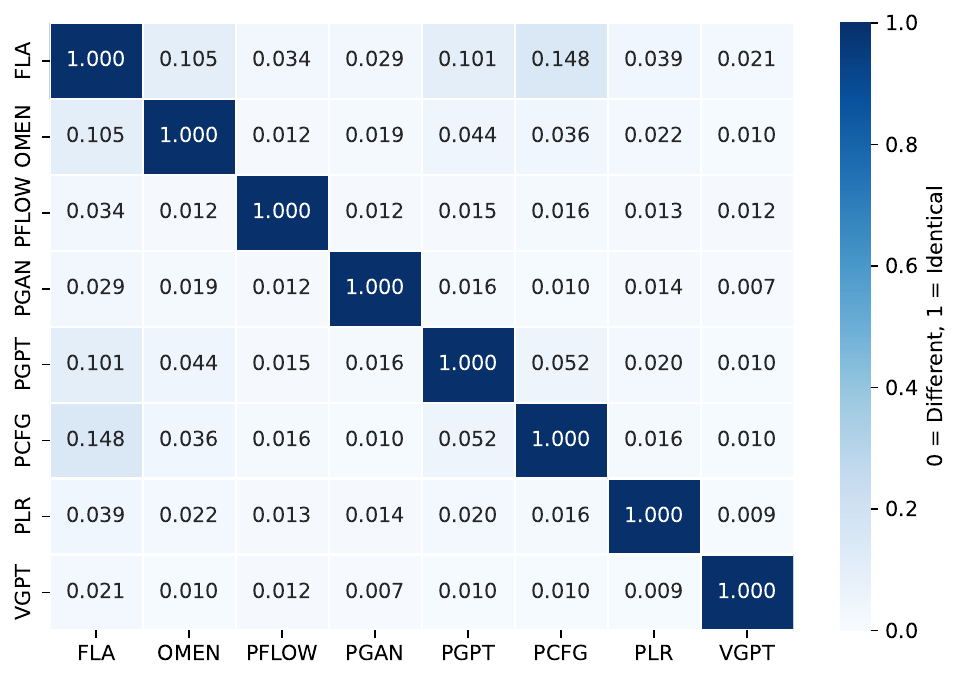}
        \label{jaccard}
    } 
    \subfloat[Mergeability Index]{\includegraphics[width=0.465\columnwidth]{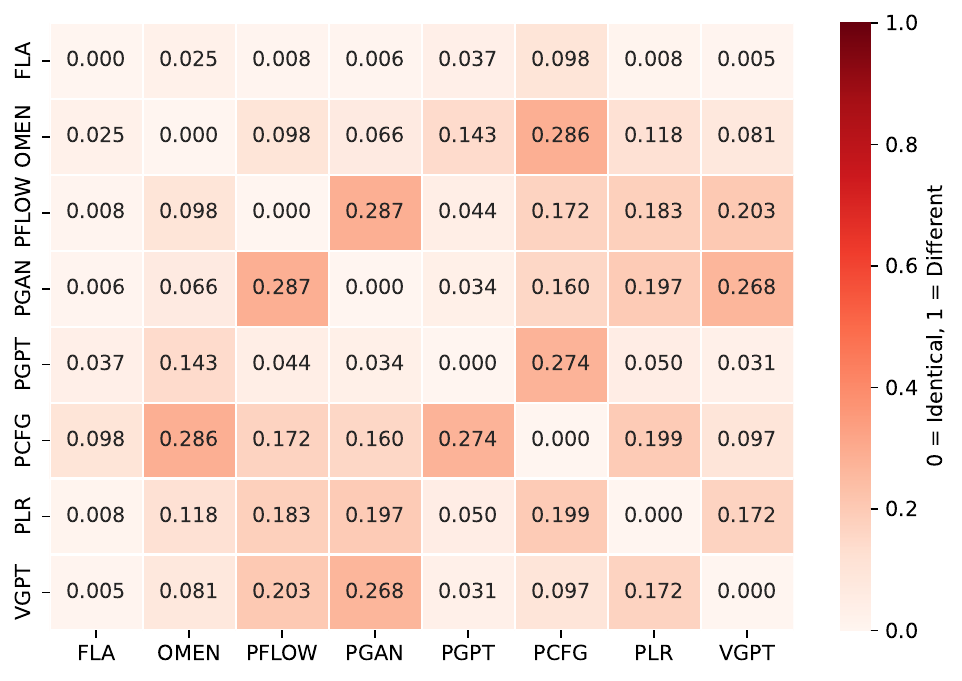}
      \label{mergeability} 
    } 
    \caption{Heatmaps of the Jaccard Index (a) and Mergeability index (b).}
    \label{fig:rq6-1}
\end{figure}

We investigate the first aspect of this RQ utilizing two primary metrics: the Jaccard Index and the Mergeability Index. Their mathematical definitions are provided in Appendix~\ref{ref:appendix_rqs}, in Equations~\ref{equation:jaccard_idx_avg} and~\ref{equation:mergeability}, respectively. 
The Jaccard Index provides a quantitative measure of the overlap between two sets of generated passwords, with a value closer to 1 indicating that the two models produce similar distributions, and a value closer to 0 indicating that they generate distinct sets of passwords. As illustrated in Figure~\ref{jaccard}, FLA-PassGPT, \rev{FLA-PCFG, and FLA-OMEN pairs are} the only combination with a Jaccard Index greater than $0.1$, suggesting that \rev{FLA shares} some similarities \rev{with these models in} password generation patterns. All other pairs exhibit much lower Jaccard values. Interestingly, despite PassGAN and PLR-GAN both being based on GANs, their Jaccard Index is quite low. Similarly, VGPT2-PassGPT also shows a low Jaccard Index. These results highlight that, even within the same architectural family, models may produce highly distinct password distributions, suggesting that combining multiple models could be beneficial for maximizing coverage of the password space.

While the Jaccard Index provides insights into the overlap of generated passwords, it does not account for their effectiveness in terms of actual matches with real-world data. To address this, we introduce the Mergeability Index, a complementary metric that measures the benefits of combining the output of two models based on their successful guesses. It captures the performance improvement achieved by merging the guesses relative to the best-performing model. A Mergeability Index close to 0 indicates that the models guess mostly the same set of passwords, while a value close to 1 indicates largely distinct matching password sets.
The results are shown in Figure~\ref{mergeability}. Interestingly, FLA exhibits the lowest mergeability across all models, indicating that its guessed passwords are generally a superset of those generated by the other models and that it has already captured a significant portion of the password space. \rev{Indeed, the only models that provide a non-negligible improvement to FLA are PCFG and, to a lesser extent, PassGPT and OMEN.}
In contrast, \rev{the other seven models} show high \rev{mergeability between them}, with their varying combinations improving successful guesses by \rev{up} to 30\%. This suggests that these models guess highly distinct sets of passwords, meaning they each \rev{capture} different aspects of the password space. 

\subsubsection{\textbf{RQ6.2 - Can We Combine Models to Maximize Effectiveness?}}

\begin{figure}[t]
    \centering
    \includegraphics[width=0.67\columnwidth]{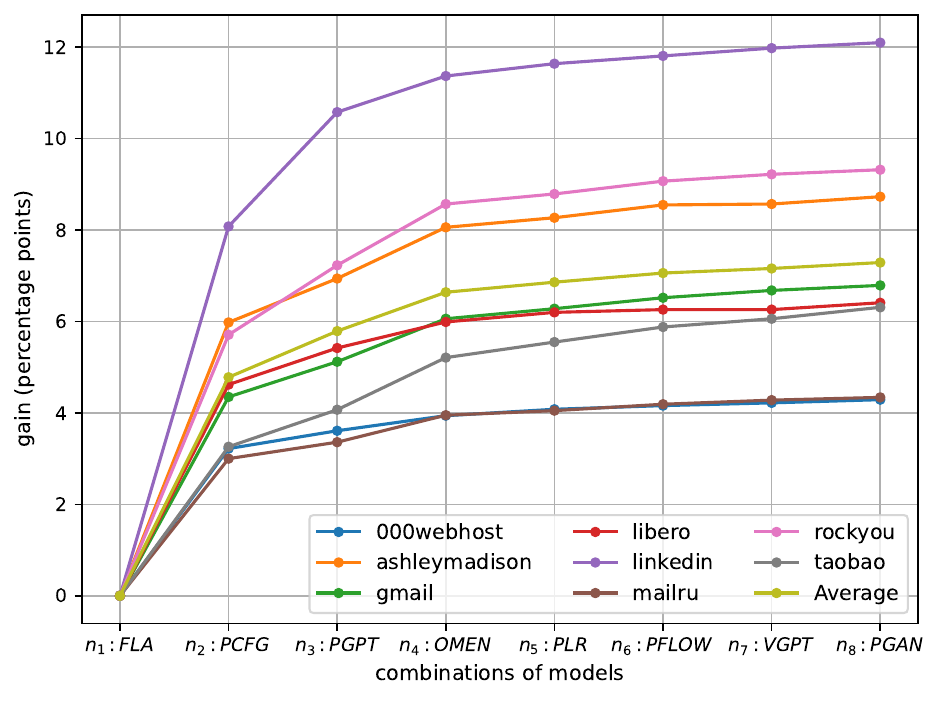}
    \caption{Multi-model guessing performance. Expressed as percentage points increase of guessed passwords relative to FLA baseline.}
    \label{fig:rq6-2}
\end{figure}

We assess the potential of a multi-model attack by evaluating the performance gain achieved through the combination of the generated password sets from $n$ models. Our results, presented in Figure~\ref{fig:rq6-2}, show the relative gain with respect to the best-performing single model, computed as percentage points on the test set. We identify the following optimal model combinations for different values of \rev{$n$: $n_1= \{FLA\}$, $n_2=n_1 + PCFG$, $n_3=n_2 + PassGPT$, $n_4=n_3 + OMEN$, $n_5=n_4 +PLR$-$GAN$, $n_6=n_5+PassFlow$, $n_7=n_6+VGPT2$, and $n_8=n_7+PassGAN$}.
In \rev{some datasets}, combining models yields significant gains over FLA \rev{alone, around 12\% for RockYou, and around 9\% for LinkedIn and Ashley Madison, mainly thanks to PCFG and} PassGPT. Remarkably, combining PCFG and FLA provides a larger average improvement than adding the remaining \rev{six} models combined. These results mirror those presented in Figure~\ref{mergeability}, indicating that FLA, \rev{PCFG}, and PassGPT capture somewhat complementary parts of the password space. \rev{In datasets like} 000webhost and Mailru, FLA \rev{and PCFG merged perform} nearly as well as the combination of all models, with \rev{less than a $0.5$} percentage point improvement \rev{when adding the other seven models}.

\subsection{\textbf{RQ7 - Do Models Truly Capture the Characteristics of Human-Like Passwords?}}\label{sec:experiments-rq7}
We assess the similarity between the generated password distribution and that of human-created passwords using multiple metrics and an analysis of their length distributions.

\subsubsection{\textbf{Metrics-Based Evaluation}}
We identified four metrics to evaluate the human-likeness of the generated passwords, each capturing different aspects and characteristics:

\subsubsection*{\textbf{CNN Divergence~\cite{cnndivergence}}} CNN Divergence utilizes a critic network trained to differentiate between real and synthetic data, with the loss serving as an estimator of their divergence. However, its reliability is influenced by the dataset size, and it may show biases when used to evaluate models trained with similar objectives, like GANs.

\subsubsection*{\textbf{IMD~\cite{imd}}} IMD is an intrinsic, multi-scale metric that compares the data manifolds of real and generated samples applicable across diverse domains. However, its effectiveness is highly sensitive to the choice of feature representation, which can introduce bias into the distance estimation.

\subsubsection*{\textbf{$\alpha$-Precision $\beta$-Recall Authenticity~\cite{precisionrecall}}} This metric characterizes distributions along three key dimensions: fidelity (precision) and diversity (recall) of the generated data, and the generalization (authentication) capabilities of the models. It is effective in identifying different types of mode failures, making it a versatile tool applicable across various domains. However, as it relies on a neural network to embed data into a hypersphere, it may introduce bias through the learned latent representation, potentially affecting the estimation of the radius used to distinguish inliers from outliers.

\subsubsection*{\textbf{MTopDiv~\cite{mtopdiv}}} MTopDiv measures topological discrepancies between real and synthetic distributions at multiple scales, effectively detecting mode failures, without relying on pre-trained networks.\\

\begin{table}[t]
    \centering
    \caption{Distance between human- and generative model-created passwords.}    
    \renewcommand{\arraystretch}{0.2}
    \resizebox{1\columnwidth}{!}{
    \begin{tabular}{m{1.2cm}<{\centering}
    m{1.6cm} m{1.6cm} m{1.6cm} m{1.6cm} m{1.6cm} m{1.6cm} }
        \toprule
        \raisebox{-0.22em}{\textbf{Models}} & \raisebox{-0.22em}{\textbf{CNN Div}}  & \raisebox{-0.22em}{\textbf{$\alpha-$Precision}} & \raisebox{-0.22em}{\textbf{$\beta-$Recall}} & \raisebox{-0.22em}{\textbf{Auth}} & \raisebox{-0.22em}{\textbf{IMD}} & \raisebox{-0.22em}{\textbf{MTopDiv}} \\
        \midrule
        \raisebox{-0.25em}{FLA} & \gradientbar{12} & \gradientbar{-15} & \gradientbar{-1} & \gradientbar{31} & \gradientbar{172} & \gradientbar{0}\\
        \raisebox{-0.25em}{OMEN} & \gradientbar{51} & \gradientbar{59} & \gradientbar{32} & \gradientbar{40} & \gradientbar{52} & \gradientbar{8}\\
        \raisebox{-0.25em}{PassFlow} & \gradientbar{56} & \gradientbar{61} & \gradientbar{52} & \gradientbar{16} & \gradientbar{200} & \gradientbar{36}\\
        \raisebox{-0.25em}{PassGAN} & \gradientbar{16} & \gradientbar{19} & \gradientbar{4} & \gradientbar{14} & \gradientbar{65} & \gradientbar{1}\\
        \raisebox{-0.25em}{PassGPT} & \gradientbar{2} & \gradientbar{3} & \gradientbar{1} & \gradientbar{6} & \gradientbar{0} & \gradientbar{0}\\
        \raisebox{-0.25em}{PCFG} & \gradientbar{19} & \gradientbar{-4} & \gradientbar{3} & \gradientbar{20} & \gradientbar{67} & \gradientbar{2}\\
        \raisebox{-0.25em}{PLR-GAN} & \gradientbar{6} & \gradientbar{-4} & \gradientbar{3} & \gradientbar{11} & \gradientbar{3} & \gradientbar{0}\\
        \raisebox{-0.25em}{VGPT2} & \gradientbar{29} & \gradientbar{53} & \gradientbar{34} & \gradientbar{4} & \gradientbar{135} & \gradientbar{12}\\
        \bottomrule
    \end{tabular}
    }
    \label{table:rq7}
\end{table}

To contextualize the metric values, we define two baselines: (1) a soft lower bound (optimal case), defined as the metric value between test and training passwords; (2) a soft upper bound (worst case), defined as the metric value between randomly generated and test passwords. For each model, we normalize the metric value between lower ($0\%$) and upper ($100\%$) bounds, as displayed in Table~\ref{table:rq7}. Values close to zero indicate closer alignment between the generated and human-created password distributions. GAN-based models demonstrate generally good performance across all metrics, with PLR consistently achieving lower values than PassGAN. PassGPT generates passwords that closely resemble human-created ones, likely due to its underlying GPT2 architecture, which is well-suited for modeling textual distributions. \rev{OMEN's passwords lie between human and random, achieving scores around 40-60\% for most metrics. PCFG struggles only on IMD, with all other indicators tending towards human values.} In contrast, PassFlow demonstrates the weakest performance, particularly in the IMD metric, where its score exceeds the soft upper bound by a factor of two. We attribute this behavior to PassFlow's Gaussian Smoothing (GS) strategy, which introduces random perturbations in the generation process. While GS enhances uniqueness~\cite{passflow}, it artificially distorts the distribution of generated passwords, ultimately reducing their quality. VGPT2 and FLA exhibit similar trends, performing strongly across most metrics except for IMD, where elevated values across most models suggest that they all struggle to fully capture certain characteristics of human passwords. Conversely, MTopDiv consistently yields low values, indicating that the models effectively avoid common mode failures, such as dropping, collapse, and invention.

\begin{figure}[t!]
    \centering
    \includegraphics[width=0.67\columnwidth]{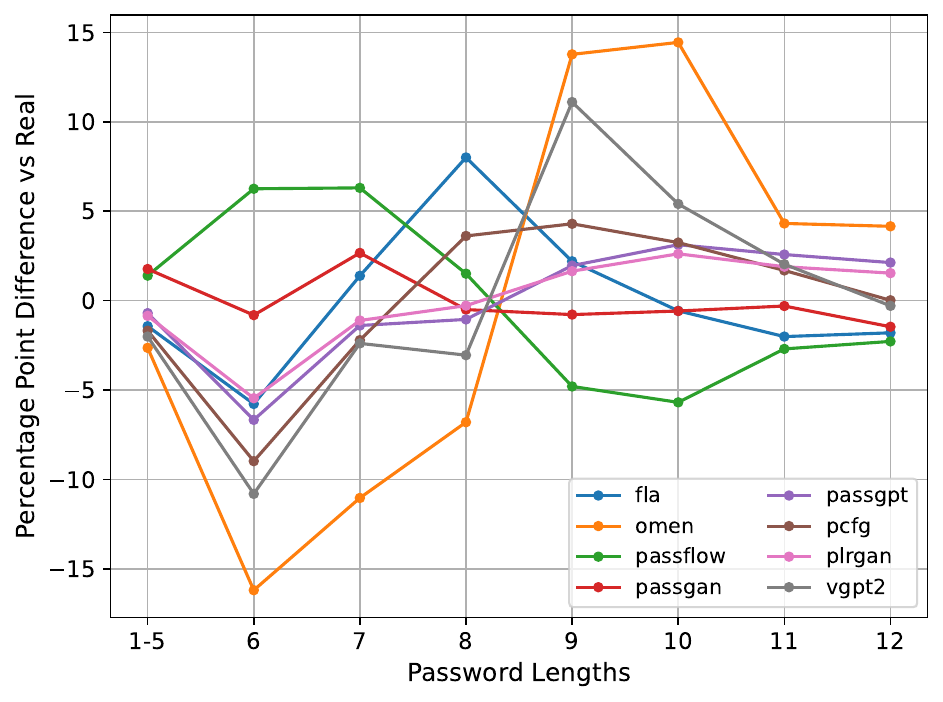}
    \caption{Password length distributions.}
    \label{fig:rq7-length-distribution}
\end{figure}

\subsubsection{\textbf{Length Distribution in Generated Passwords}}
Figure~\ref{fig:rq7-length-distribution} compares the length distribution of generated passwords with that of real passwords. Among the models, \rev{OMEN}, VGPT2, FLA, PassFlow, \rev{and to a lesser extent PCFG}, show the largest divergence from the real distribution: PassFlow tends to generate shorter passwords, FLA overestimates the probability of 8-character passwords, while \rev{OMEN, VGPT2, and PCFG} of longer ones. In contrast, PassGAN, PLR-GAN, and PassGPT produce distributions that more closely align with real data, reinforcing previous observations. 

\subsubsection{\textbf{Why Does IMD Yield High Values?}}
Figure~\ref{fig:rq7-length-distribution} reveals a clear correlation between length distribution and IMD values. PassFlow, which exhibits \rev{one of} the largest deviations from the real distribution, also records the highest IMD score. A similar trend is observed for FLA, VGPT2, \rev{OMEN, and PCFG}. Conversely, models such as PassGAN, PLR-GAN, and PassGPT, which more closely replicate the real-world length distribution, achieve lower IMD scores. Interestingly, Table~\ref{table:rq7} shows that models such as PassFlow and FLA yield higher IMD scores than those associated with the random-password soft upper bound. We posit that this counterintuitive result stems from the uniform length distribution of random passwords. Although this distribution diverges significantly from that of real passwords, it nonetheless includes longer passwords that PassFlow and FLA struggle to generate. These observations suggest that IMD is highly sensitive to length distribution, favoring models that replicate it more accurately. 

\begin{table}[t!]
    \centering
    \caption{Comparison summary of the selected approaches.}
    \label{tab:benchmarking}
    \resizebox{\columnwidth}{!}{
        \begin{tabular}{|l|c|c|c|c|c|c|c|c|}
    \toprule
    \textbf{Metric} & \textbf{FLA} & \textbf{OMEN} & \textbf{PFLOW} & \textbf{PGAN} & \textbf{PGPT} & \textbf{PCFG} & \textbf{PLR} & \textbf{VGPT} \\
    \midrule
    \multicolumn{9}{|l|}{\textbf{Overall Performance}} \\ 
    \quad Scenario: In-Distribution $\uparrow$ & 49.21 & 29.07 & 12.76 & 8.85 & 37.21 & 37.97 & 14.88 & 11.79 \\ 
    \quad Scenario: Cross-Community $\uparrow$ & 27.72 & 12.23 & 7.13 & 3.62 & 17.72 & 25.13 & 6.16 & 4.76 \\
    \quad Scenario: Cross-Culture $\uparrow$ & 23.27 & 13.55 & 16.17 & 6.52 & 14.24 & 12.11 & 9.70 & 7.01 \\
    \midrule
    \multicolumn{9}{|l|}{\textbf{Performance by Frequency}} \\ 
    \quad Frequency: Common (Top 5\%) $\uparrow$ & 83.37 & 66.65 & 31.20 & 27.87 & 68.20 & 77.39 & 36.86 & 32.83 \\
    \quad Frequency: Common (Top 10\%) $\uparrow$ & 76.60 & 57.64 & 25.73 & 22.73 & 63.45 & 67.50 & 30.63 & 27.23 \\
    \quad Frequency: Rare (Bottom 90\%) $\uparrow$ & 47.03 & 27.40 & 12.28 & 8.75 & 34.51 & 33.78 & 13.61 & 10.83 \\
    \midrule
    \multicolumn{9}{|l|}{\textbf{Performance by Length}} \\ 
    \quad Length: Short (4-7 Chars) $\uparrow$ & 72.97 & 32.42 & 28.16 & 19.91 & 55.62 & 49.60& 32.96 & 29.65 \\
    \quad Length: Medium (8-10 Chars) $\uparrow$ & 36.38 & 20.33 & 1.33 & 2.48 & 26.93 & 33.44 & 4.47 & 2.16 \\
    \quad Length: Long (11-12 Chars) $\uparrow$ & 11.60& 2.64 & 0.00 & 0.07 & 12.58 & 15.56 & 0.24 & 0.05 \\
    \midrule
    \multicolumn{9}{|l|}{\textbf{Performance by Pattern}} \\
    \quad Pattern: Simple (1 Char Class) $\uparrow$ & 53.52 & 33.97 & 16.22 & 14.06 & 41.16 & 24.42 & 21.71 & 17.85 \\
    \quad Pattern: Moderate (2 Char Classes) $\uparrow$ & 38.94 & 15.69 & 4.71 & 3.46 & 29.56 & 42.40 & 7.15 & 5.26 \\
    \quad Pattern: Complex (3 Char Classes) $\uparrow$ & 14.20 & 2.92 & 0.05 & 0.08 & 11.19 & 12.11 & 0.27 & 0.33 \\
    \midrule
    \multicolumn{9}{|l|}{\textbf{Generalizability}} \\ 
    \quad Train Set Size Sensitivity (\%) $\downarrow$ & 7.58 & 1.15 & 1.90 & 12.56 & 13.16 & 9.96 & 3.54 & 35.49 \\
    \quad Cross-Community Loss (\%) $\downarrow$ & 30.13 & 20.54 & 6.84 & 24.16 & 42.15 & 28.58 & 32.42 & 32.49 \\ 
    \quad Cross-Culture Loss (\%) $\downarrow$ & 58.53 & 62.13 & 14.58 & 48.67 & 66.15 & 69.31 & 49.27 & 57.11 \\ 
    \midrule
    \multicolumn{9}{|l|}{\textbf{Generated Password Quality}} \\
    \quad Uniqueness (\%) $\uparrow$ & 100.00 & 100.00 & 90.10 & 54.59 & 73.22 & 99.89 & 71.13 & 91.41 \\
    \quad Humanness Distance (\%) $\downarrow$ & 38.50 & 40.33 & 70.16 & 19.83 & 2.40 & 19.16 & 4.80 & 44.00 \\
    \bottomrule
    \end{tabular}
    }
\end{table}

\subsection{\textbf{Resource Consumption}}
\rev{ 
Experiments were conducted on an Ubuntu 24.04.2 LTS server equipped with 2x AMD EPYC 9354 (32 cores per CPU, 2 threads per core), 503 GB of RAM, and 5x NVIDIA RTX A6000 with 48 GB of VRAM each. In Table~\ref{tab:resource} we report resource consumption during training, computed over a training dataset of 1 million passwords with a maximum length of 12, and resource consumption during sampling, computed when generating $5 \times 10^8$ passwords. Except for GANs, the reported training time multiplied by $x$ denotes the expected time to train on $x$ million passwords. By design, GAN-based approaches are independent of dataset size, as training stops after a fixed number of generator iterations; thus, the reported total training time reflects the expected time to train such models. 

OMEN and PCFG exhibit clear advantages, as they are the fastest and most memory-efficient methods for training and sampling. Regarding DL-based approaches, the resource consumption highly depends on the model size, architecture, and guessing strategy. Although the training and sampling times are higher than those of traditional tools, they are not prohibitively so; the primary concern lies in memory usage. PassGPT, PassFlow, and partially FLA are computationally demanding, making them practical only in unconstrained environments (e.g., clusters). During sampling, FLA requires $\sim 28$ GB of RAM to complete the algorithm described in Appendix~\ref{sec:appendix-models}. Similarly, PassFlow's GS technique requires storing in memory all unique passwords generated for the whole sampling process, corresponding to $\sim 43$ GB of RAM when generating $5 \times 10^8$ passwords. Instead, PassGPT requires at least a GPU equipped with $>$24 GB of VRAM during training, and $\sim 51$ GB of RAM during sampling. The remaining three models: PassGAN, PLR-GAN, and VGPT2 are more resource-efficient, requiring at most 2 GB of RAM during training and 4 GB of RAM during sampling, making them suitable for deployment in various environments. 

Regarding model size, for DL-based approaches, we report the size of the model itself. In contrast, for OMEN and PCFG, the reported size refers to the average disk space required to store the generated rules. FLA is by far the largest model among the others, weighing 391 MB. The transformer-based models, PassGPT and VGPT2, follow, with sizes slightly above 200 MB. All the other approaches can be considered lightweight, occupying less than 100 MB.}

\begin{table}[t!]
    \centering
    \caption{\rev{Time and memory consumption of the selected approaches. Training stats follow the format h:m:s.} }
    \label{tab:resource}
    \renewcommand{\arraystretch}{1.10}
    \begin{adjustbox}{width=0.98\columnwidth, center}
    \begin{tabular}{l|l l l l|l l l|l}
        \toprule
        \multicolumn{1}{c}{\textbf{}} & \multicolumn{4}{c}{\textbf{\rev{Training Stats}}} & \multicolumn{3}{c}{\textbf{\rev{Sampling Stats}}} & \multicolumn{1}{c}{\textbf{\rev{General}}} \\
        \cmidrule(l){2-5} \cmidrule(lr){6-8} \cmidrule(l){9-9}
        {\textbf{\rev{Model}}} & {\textbf{\rev{1 Epoch}}} & {\textbf{\rev{Total Time}}} & {\textbf{\rev{RAM Peak}}} & {\textbf{\rev{VRAM Peak}}} & {\textbf{\rev{PW/s}}} &{\textbf{\rev{RAM Peak}}} &{\textbf{\rev{VRAM Peak}}}  &{\textbf{\rev{Model Size}}} \\
        \midrule
        FLA & 00:06:38 & 02:16:40 & 1.31GB & 10.07GB & 9926 & 28.74GB & 4.97GB & 391.2MB\\
        OMEN & 00:00:39 & 00:00:39 & 0.07GB & N/A & 653594 & 0.77GB & N/A & 81.64MB\\
        PassFlow & 00:00:45 & 02:30:00 & 1.95GB & 0.47GB & 17538 & 43.63GB & 0.43GB & 59.2MB \\
        PassGAN & 00:05:58 & 12:43:44 & 2.13GB & 0.43GB & 77483 & 4.38GB & 0.41GB & 22.4MB\\
        PassGPT & 00:06:52 & 00:19:36 & 2.65GB & 23.63GB & 3672 & 51.13GB & 5.19G & 230.2MB \\
        PCFG & 00:00:40 & 00:00:40 & 0.07GB & N/A & 370370 & 0.67GB & N/A & 84.29MB\\
        PLR-GAN & 00:09:00 & 38:24:00 & 2.16GB & 0.40GB & 62617 & 4.89GB & 0.42GB & 11.2MB\\
        VGPT2 & 00:03:26 & 01:01:20 & 1.90GB & 1.31GB & 4374 & 4.03GB & 1.78GB & 206.3MB\\
        \bottomrule
    \end{tabular}
    \end{adjustbox}
\end{table}

\subsection{\textbf{General Benchmarking}}\label{sec:experiments-benchmark}
This section presents a comprehensive benchmark of the selected generative models, leveraging prior evaluations to facilitate direct comparisons across key metrics: performance, generalizability, and generated password quality. Table~\ref{tab:benchmarking} summarizes all aggregated results, reflecting the trends observed in the preceding sections. Regarding performance metrics, computed as the weighted average percentage of guessed passwords, FLA \rev{generally} achieves the best results across tasks, with PassGPT, \rev{PCFG, and OMEN, as its main competitors}. The remaining models exhibit greater task dependency: for instance, PLR-GAN leads in in-distribution scenarios, while PassFlow excels in cross-community and cross-culture settings. However, when evaluating generalizability, the overall ranking shifts. We consider three metrics: sensitivity to training set size, measured using the coefficient of variation, and performance loss in cross-community and cross-culture scenarios, quantified as weighted average loss relative to the in-distribution setting. PassFlow consistently emerges as the most robust model across generalization tasks. Surprisingly, FLA, \rev{PCFG, OMEN}, and PassGPT exhibit the highest loss in the cross-cultural setting, with PassGPT \rev{, OMEN, and PCFG} additionally demonstrating the lowest robustness in the cross-community scenario, whereas FLA maintains an average value. Finally, concerning the quality metrics, we analyzed both uniqueness and humanness of generated passwords. While not previously emphasized in our analysis, uniqueness is a commonly adopted metric for evaluating the diversity of generated passwords. While FLA's generation approach guarantees 100\% uniqueness by design, PassFlow, leveraging the GS technique, and VGPT2 produce around 90\% unique passwords. PLR-GAN and PassGPT achieve slightly above 70\% uniqueness, and PassGAN demonstrates the lowest uniqueness at 54.59\%. Regarding humanness distance, measured as the average percentage distance between generated and real passwords, results align with our previous observations: models such as PassGPT and PLR-GAN generate passwords closely resembling human-created ones, making them highly suitable for applications like honeywords\rev{~\cite{juels2013honeywords}}, where it is important to deceive adversaries using realistic passwords, and as potential sources of synthetic data for future research, thereby addressing challenges with real-world password datasets (e.g., difficult to obtain, outdated, ethical concerns).
\section{Insights and Lessons Learned}\label{sec:lesson_learned}
This section outlines key insights obtained in our study.

\subsubsection*{\textbf{\rev{DL-based Autoregressive} Models Perform Best}}
FLA and PassGPT, based on LSTM and GPT2, consistently outperform other \rev{generative approaches}, underscoring the advantages of \rev{deep learning autoregressive} models in capturing dependencies and generating more accurate, complex guesses.

\subsubsection*{\textbf{Generative Approaches Surpass Traditional Rule-Based Tools}}
Early generative approaches---such as PassGAN, PassFlow, and PLR-GAN--- struggle to match the performance of traditional rule-based tools. However, the emergence of transformer-based architectures has shifted the research landscape toward models capable of consistently outperforming them. PassGPT, the most recent approach evaluated in our study, generally surpasses JtR and Hashcat. As research progresses, more advanced LLMs are likely to further improve password-guessing effectiveness, whereas traditional tools have likely reached their performance ceiling.

\rev{\subsubsection*{\textbf{Traditional ML-Based Approaches Remain Competitive}} Traditional ML-based approaches remain widely competitive, with PCFG and OMEN generally ranking between 2nd and 4th across each benchmarked metric. Furthermore, they also demonstrate strong generalizability and, for PCFG, good performance on longer passwords and complex patterns.}

\subsubsection*{\textbf{Rare Does Not Mean Hard to Guess}}
Rare passwords, such as those appearing only once in the dataset, are not necessarily hard to guess, especially if they are semantically similar to common passwords.

\subsubsection*{\textbf{Stricter Policies Mean Safer Passwords}}
Our analysis reveals that each dataset exhibits a distinct distribution, with some being significantly easier to guess than others. Notably, 000webhost and LinkedIn---both of which follow stricter password policies regarding length and complexity---emerge as the most challenging datasets. These findings reinforce the conventional understanding that even modestly strict password policies enhance security---an effect that holds true even in the context of generative models.

\subsubsection*{\textbf{Guessing Complex/Long Passwords Remains Challenging}} Four out of \rev{eigth} generative models struggle to guess complex passwords---those containing special characters or multiple character classes---and longer passwords exceeding eight characters. Only FLA, PassGPT, \rev{and the ML-based tools OMEN and PCFG} achieve a non-negligible success rate in these cases, highlighting the persistent challenge posed by long and complex passwords and revealing model limitations. 

\rev{\subsubsection*{\textbf{Models Reach a Hard Ceiling Even with Larger Datasets}}
Based on the insights obtained in RQ3, we expect larger datasets to further improve performance for certain models, including PassGPT, VGPT2, and PCFG. Being an advanced autoregressive model, PassGPT is expected to scale particularly well, following trends observed in NLP~\cite{kaplan2020scaling, henighan2020scaling}. However, all models are likely to reach a hard ceiling as they approach the limits of modeling complex/infrequent passwords that lie on the extremities of the learned distribution, which cannot be fully captured even with larger datasets.}

\subsubsection*{\textbf{Models Go Beyond Memorization}} 
While performance generally declines relative to in-distribution settings, models still generalize well to unseen distributions, indicating they go beyond simple memorization. Additionally, some models exhibited greater robustness to distribution shifts, even if they performed worse under ideal, in-distribution settings.

\subsubsection*{\textbf{Models Generate and Match Distinct Passwords}} Even when trained on the same dataset, models generate and match distinct sets of passwords. This diversity enables multi-model attacks that outperform the best individual model.

\subsubsection*{\textbf{Models Generate Human-Like Passwords}}
The selected models have been shown to effectively learn to generate passwords that closely resemble human-created ones, making them valuable for a variety of tasks.

\subsubsection*{\textbf{Beyond Guessing Rate}}
Prior research primarily assesses password-guessing tools by focusing on guessing rate, thus providing a narrow view of their capabilities. We argue that the emergence of generative password-guessing requires a broader scope, including additional key metrics to assess the quality of the generated data (e.g., humanness, uniqueness) and the generalizability of the models (e.g., train set size sensitivity, cross-dataset loss). Together with guessing performance, these metrics provide a more comprehensive evaluation and characterization of generative models.

\rev{\subsubsection*{\textbf{On Practical Implications}}
To improve password security, real-world systems usually rely on password policies and complementary mechanisms such as honeywords~\cite{juels2013honeywords}, password strength meters~\cite{bergadano1998high, bishop1995improving}, and password blocklists. Our RQ7 findings show that some approaches, such as PassGPT, can generate exceptionally human-like passwords. This ability is directly relevant for strengthening the generation of realistic honeywords~\cite{dionysiou2021honeygen} and improving existing password blocklists with easily-guessable passwords. Additionally, effective generative models can also be leveraged to improve the effectiveness of password strength meters~\cite{liu2023confident}.}
\section{Conclusion}
This work presented \name, a framework designed to comprehensively evaluate password-guessing approaches. It includes standardized guidelines and advanced testing scenarios to ensure fair, in-depth comparisons and reveal the strengths and limitations of different models. By analyzing eight real-world password leaks and thoroughly evaluating various password-guessing approaches, we addressed seven key research questions, offering insights into the factors influencing user password choices and the current state of generative password-guessing research. We believe future research could greatly benefit from \name, as it can accelerate the development of new approaches beyond password-guessing, such as enhancing password security mechanisms like honeywords and password strength meters. \name is intended to foster academic advancements in password security, not to facilitate or promote malicious activity.
\rev{\section*{Acknowledgments}
This work was supported by (1) Project SERICS under the National Recovery and Resilience Plan by the Minister of University and Research (NRRP MUR) Program funded by the European Union-NextGenerationEU under Grant PE00000014, (2) the NRRP MUR Program funded by the European Union-NextGenerationEU, Mission 4, Component 1, CUP B53C24002200004, and (3) the "Progetto di Ricerca Grandi 2022, Sapienza University of Rome".}

\bibliographystyle{IEEEtranS}
\bibliography{bibliography}

\appendices{}

\begin{table*}[t!]
    \centering
    \caption{Distribution of password patterns from r1 to r19. }
    \label{tab:pattern_distribution}
    \renewcommand{\arraystretch}{1.15}

    \begin{adjustbox}{width=0.9\textwidth, center}
    \begin{tabular}{l *{19}{c}}
        \toprule
        \textbf{Dataset} & {\textbf{r1}} & {\textbf{r2}} & {\textbf{r3}} & {\textbf{r4}} & {\textbf{r5}} & {\textbf{r6}} & {\textbf{r7}} & {\textbf{r8}} & {\textbf{r9}} & {\textbf{r10}} & {\textbf{r11}} & {\textbf{r12}} & {\textbf{r13}} & {\textbf{r14}} & {\textbf{r15}} & {\textbf{r16}} & {\textbf{r17}} & {\textbf{r18}} & {\textbf{r19}} \\
        \midrule
        rockyou  & 44.28 & 41.89 & 1.51 & 15.93 & 0.02 & 36.26 & 1.64 & 0.15 & 1.67 & 31.64 & 1.11 & 0.67 & 0.10 & 16.54 & 0.04 & 0.17 & 0.07 & 0.47 & 9.38 \\
        linkedin & 20.38 & 18.02 & 0.78 & 19.59 & 0.01 & 53.22 & 1.28 & 0.16 & 5.36 & 43.10 & 1.97 & 0.66 & 0.25 & 21.23 & 0.07 & 0.35 & 0.31 & 0.82 & 10.02 \\
        mailru   & 27.19 & 24.37 & 0.27 & 18.54 & 0.00 & 52.19 & 0.62 & 0.35 & 1.12 & 27.00 & 0.14 & 0.37 & 0.16 & 21.23 & 0.03 & 0.04 & 0.06 & 0.03 & 5.75 \\
        000web   & 0.42  & 0.18  & 0.01 & 0.04  & 0.02 & 92.89 & 1.20 & 0.40 & 4.97 & 69.11 & 1.77 & 0.66 & 0.50 & 3.31  & 0.00 & 0.02 & 0.02 & 0.43 & 13.10 \\
        taobao   & 15.70 & 15.42 & 0.13 & 27.90 & 0.01 & 55.89 & 0.09 & 0.10 & 0.31 & 50.18 & 0.06 & 0.04 & 0.08 & 28.46 & 0.00 & 0.01 & 0.01 & 0.01 & 9.92 \\
        gmail    & 39.78 & 39.78 & 0.00 & 15.70 & 0.02 & 42.38 & 0.86 & 0.14 & 1.13 & 33.36 & 0.53 & 0.58 & 0.09 & 16.93 & 0.12 & 0.07 & 0.17 & 0.00 & 9.05 \\
        ashleym  & 35.04 & 33.18 & 0.98 & 12.32 & 0.00 & 52.24 & 0.12 & 0.02 & 0.27 & 41.79 & 0.11 & 1.05 & 0.01 & 13.43 & 0.00 & 0.01 & 0.01 & 0.05 & 10.11 \\
        libero   & 42.04 & 39.01 & 1.86 & 13.10 & 0.00 & 42.41 & 0.63 & 0.08 & 1.69 & 34.06 & 0.75 & 0.24 & 0.09 & 15.94 & 0.02 & 0.09 & 0.05 & 0.24 & 6.95 \\
        \bottomrule
    \end{tabular}
    \end{adjustbox}
\end{table*}

\begin{table*}[t!]
    \centering
    \caption{Top 10 passwords for each dataset. The password at the 8th position in 000webhost (indicated as ``*'') is ``YfDbUfNjH10305070'': the letter portion of the password can be mapped to a Russian word meaning ``Navigator''. The reasons for its unexpected popularity remain unclear~\cite{wang2016implications}.}
    \label{tab:top10}
    \renewcommand{\arraystretch}{1.15}

    \begin{adjustbox}{width=0.9\textwidth, center}
    \begin{tabular}{l *{10}{l}}
        \toprule
        \textbf{Dataset} & \textbf{1} & \textbf{2} & \textbf{3} & \textbf{4} & \textbf{5} & \textbf{6} & \textbf{7} & \textbf{8} & \textbf{9} & \textbf{10} \\
        \midrule
        Rockyou     & 123456      & 12345       & 123456789   & password    & iloveyou      & princess    & rockyou     & 1234567     & 12345678    & abc123 \\
        Linkedin    & linkedin    & 123456      & 123456789   & abc123      & idontknow     & ilovelinkedin & Godisgood   & jaimatadi   & linkedin1   & iloveindia \\
        Mailru      & qwerty      & qwertyuiop  & 123456      & qwe123      & qweqwe        & klaster     & 1qaz2wsx    & 1q2w3e4r    & qazwsx      & 1q2w3e \\
        000webhost  & abc123      & 123456a     & 12qw23we    & 123abc      & a123456       & 123qwe      & secret666   & * & asd123      & qwerty123 \\
        Taobao      & 123456      & 111111      & 123456789   & 123123      & 000000        & 5201314     & wangyut2    & 123         & 123321      & 12345678 \\
        Gmail       & 123456      & password    & 123456789   & 12345       & qwerty        & 12345678    & 111111      & abc123      & 123123      & 1234567 \\
        Ashley M.   & eatpussy    & opensaysme  & christina   & longing     & nastygirl     & steve       & 11inches    & 2ofus       & 69sex       & 99wmp \\
        Libero      & 123456      & popopo90    & francesco   & 123456789   & 12345678      & napoli      & alessandro  & amoremio    & andrea      & francesca \\
        \bottomrule
    \end{tabular}
    \end{adjustbox}
\end{table*}

\section*{Ethical Considerations}\label{sec:ethical}
In line with prior research~\cite{fla, plr-gan, passgpt, xu2023improving}, we consider the use of leaked datasets to be ethical, as: (1) they are publicly available, (2) their usage does not cause additional harm, (3) we do not use any additional sensitive information, such as email addresses, phone numbers, or usernames, that could link specific passwords to individual users, and (4) such data are essential for advancing research. We discourage any usage of \name for illegal or unethical purposes. The framework is developed exclusively for academic research to drive advancements in password security.

\section{\textbf{Models and Tools Implementation}} \label{sec:appendix-models}
This appendix provides additional implementation details for each selected model and presents a categorization of each approach in Table~\ref{tab:categorization}. \rev{Regarding deep generative models, we standardized dependencies by porting them to PyTorch 2.6.0, as some were originally implemented in TensorFlow or outdated PyTorch versions. To provide a unified comparison framework,  we designed an interface and adapted each model accordingly. Future research can easily integrate new models by subclassing the base model class and implementing the required abstract methods.} Unless otherwise stated, all training parameters are the same as those used in the original works. \rev{We defined a validation procedure for our implementation and assessed its accuracy by comparing our results with those reported in the original papers, observing negligible variations falling within the margin of error. The sole exception was VGPT2, for which we were unable to replicate the published results using the official implementation. }

\subsubsection*{\textbf{FLA}}
We implemented FLA following the description of the large model mentioned in~\cite{fla}. Specifically, the architecture consists of three LSTM layers with 1000 cells each, followed by two FC layers. Each checkpoint was trained for 20 epochs, processing input backwards. Unlike other generative approaches, FLA requires a probability threshold as input, filtering out passwords whose overall probability falls below it. We set this threshold to $10^{-8}$ for up to $10^6$ generated passwords, $10^{-9}$ for $10^7$, and $10^{-10}$ for $5 \times 10^8$. After generation, we sort guesses by probability in descending order and select the first $n$ ones, equivalent to finding the optimal threshold to generate exactly $n$ passwords. 
\rev{Sorting $5 \times 10^8$ passwords is highly computationally demanding; therefore, we implemented a custom min-heap in Cython, which significantly improved FLA's resource efficiency.}

\subsubsection*{\textbf{PassGAN}}
We implemented PassGAN as described in the original work~\cite{passgan}: using an IWGAN~\cite{iwgan} with both the generator and discriminator composed of 5 residual blocks~\cite{he2015deepresiduallearningimage}. We trained PassGAN's models for $200,\!000$ iterations, validating checkpoints every $10,\!000$ steps.

\subsubsection*{\textbf{PLR-GAN}}
PLR-GAN~\cite{plr-gan} is an enhanced version of PassGAN aimed at improving training instability. PLR injects low-magnitude noise into the one-hot character encodings during training, enabling the integration of deeper architectures and longer training. While the original work states that PLR can be trained up to 4 million iterations, we capped our training at $400,000$ iterations due to computational constraints and the diminishing returns observed beyond this point. All our experiments utilize the Dynamic Password Guessing strategy proposed by the authors.

\subsubsection*{\textbf{PassFlow}}
We optimized the original PassFlow implementation~\cite{passflow} to improve efficiency. We introduced an early-stopping mechanism that halts the training if the model fails to improve its performance by at least 5\% over the current best result for 10 consecutive epochs, starting from epoch 100. Otherwise, training continues for 200 epochs. All our experiments were conducted using PassFlow's Gaussian Smoothing technique. The original GS implementation adds noise to the generated passwords until a unique one is generated. However, this approach is highly inefficient on certain datasets, significantly slowing down generation (up to weeks of time for a single run). We modified the GS algorithm, introducing an early stopping mechanism if no unique password is generated after 100 iterations.

\subsubsection*{\textbf{VGPT2}}
We implemented VGPT2~\cite{vgpt2} without introducing any architectural or training modifications. \rev{We followed the official code for the implementation, unifying the testing methodology with our framework.}

\subsubsection*{\textbf{PassGPT}}
We used the original PassGPT implementation, applying minimal changes to ensure compatibility with our framework. Following the authors' recommendations~\cite{passgpt}, we trained PassGPT on a deduplicated dataset.

\subsubsection*{\textbf{OMEN}}
We used the publicly available implementation of OMEN~\cite{durmuth2015omen} from~\cite{pcfg-omen-github}. \rev{Following~\cite{durmuth2015omen},} we configured it as a 4-gram Markov model with a coverage set to 0.

\subsubsection*{\textbf{PCFG}}
We used the publicly available implementation of PCFG from~\cite{pcfg-omen-github}. We set the coverage parameter to 1, thereby relying exclusively on guesses produced by the PCFG model.

\section{\textbf{Additional Details On RQs}} \label{ref:appendix_rqs}

\subsection{\textbf{Additional Details On RQ6}}
To address RQ6, we employed two metrics: the Jaccard Index and the Mergeability Index. We now present their formal definitions:

\subsubsection*{\textbf{Jaccard Index:}} 
Given two models $m_1$ and $m_2$ and a set of datasets D = $\{D_1, D_2, ..., D_z\}$, let $f(m_i,D_j^{train},s) = P_{i,j}$ denote the set of passwords generated by model $m_i$ after being trained on dataset $D_j^{train}$, using settings $s$. We define the Jaccard Index between $m_1$ and $m_2$ as:
\begin{equation}
    J(m_1, m_2, D) = \frac{1}{|D|} \sum_{j=1}^{|D|} \frac{|(P_{1,j} \cap P_{2,j}) |}{|(P_{1,j} \cup P_{2,j})|}
\label{equation:jaccard_idx_avg}
\end{equation}
\begin{equation*}
    \text{where: } P_{i,j} = f(m_i,D_j^{train},s)
\end{equation*}

\subsubsection*{\textbf{Mergeability Index:}}
Let $G_{i,j}$ = $P_{i,j} \ \cap \ D^{test}_j$ denote the set of passwords generated by model $m_i$, after being trained on $D_j^{train}$,  that match those in the corresponding testing set $D_j^{test}$. The Mergeability Index is defined as follows:

\begin{equation}
    MI(m_1, m_2, D) = \frac{1}{|D|} \sum_{j=1}^{|D|}  \left( \frac{| (G_{1,j} \cup G_{2,j})| - G_{\text{MAX}}}{G_{\text{MAX}}} \right)
    \label{equation:mergeability}
\end{equation}
\begin{equation*}
    \text{where: }
    G_{\text{MAX}} = \max(|G_{1,j}|, |G_{2,j}|).
\end{equation*}

\subsection{\textbf{Additional Details On RQ7}}
Let $D^{train}=\bigcup_{j=1}^{|D|} D^{train}_{j}$ and $D^{test}=\bigcup_{j=1}^{|D|} D^{test}_{j}$ denote the union of all training and testing datasets, respectively. Let $R$ be the set of randomly generated passwords between 6 and 12 characters in length. The lower ($L$) and upper ($U$) baselines for a metric $dist$ are computed as follows:

\begin{equation}
    L = dist(D^{\text{test}}, D^{\text{train}}), \quad
    U = dist(D^{\text{test}}, R)
\end{equation}

Next, for each model $m_i$, let $P_{i}$ be the set of passwords generated by model $m_i$ across all datasets. The value $d_i$ for a model $m_i$ on a given metric $dist$ is then computed using the following equation:

\begin{equation}
    d_i = \frac{dist(D^{test}, P_i) - L}{U -L} * 100
\end{equation}

\subsubsection*{\textbf{CNN Divergence~\cite{cnndivergence}}} Neural networks can be used to estimate the divergence between two distributions, making them useful for evaluating generative models. The idea is to employ an independent critic network trained to distinguish between real and generated samples. After sufficient training, the critic’s loss, based on WGAN-GP~\cite{iwgan}, reflects the distance between the two distributions. Our convolutional neural network (CNN) follows the architecture and settings described in~\cite{cnndivergence}, based on the DCGAN discriminator~\cite{radford2015unsupervised}. 

\subsubsection*{\textbf{IMD~\cite{imd}}} Intrinsic Multi-scale Distance is a metric designed to compare the data manifolds of two distributions. IMD provides an intrinsic method to lower-bound the spectral Gromov-Wasserstein distance between two manifolds. Unlike other approaches that focus on extrinsic properties and are uni-scale, IMD is intrinsic, meaning it is not dependent on the transformation of the manifold and multi-scale, capturing both local and global properties.

\subsubsection*{\textbf{$\alpha$-Precision $\beta$-Recall Authenticity~\cite{precisionrecall}}} The $\alpha$-Precision $\beta$-Recall Authenticity is a three-dimensional metric, with each dimension corresponding to a distinct property:
\begin{itemize}
    \item \textbf{Fidelity ($\alpha$-Precision):}  Measures how closely the generated samples resemble the most typical fraction ($\alpha$-support) of real data.
    \item \textbf{Diversity ($\beta$-Recall):} Represents the fraction of real samples that lie within the $\beta$-support of the generated data distribution.
    \item \textbf{Generalization (Authenticity):} Reflects the model’s ability to generalize, ensuring that the output is not limited to mere copies of the training data.
\end{itemize}
Each dimension is computed after mapping generated and real data into a hypersphere using a feature embedding, where most of the data is concentrated near the center, while outliers are positioned closer to the boundaries.

\subsubsection*{\textbf{MTopDiv~\cite{mtopdiv}}} MTopDiv measures multiscale topology discrepancies between two distributions, P and Q, in a high-dimensional space. Unlike IMD, MTopDiv also considers extrinsic properties, using position and translation to capture structural differences between the distributions.

\begin{table}[t!]
    \centering
    \caption{IMD outputs as the distributions $P$ and $Q$ vary.}
    \label{tab:rq7-analysis1}
    \renewcommand{\arraystretch}{1.2}

    \begin{adjustbox}{width=0.95\columnwidth, center}
    \begin{tabular}{c c}
        \toprule
        \textbf{IMD(P,Q)} & {\textbf{Output}} \\
        \midrule
        P = 000webhost - Q = Random                     & 7.9919 \\
        P = 000webhost - Q = PassFlow 000webhost        & 47.0466 \\
        P = PassFlow 000webhost - Q = Random            & 49.0287 \\
        P = PassFlow 000w. - Q = Rand with PassFlow 000w. Length & 30.9260 \\
        \bottomrule
    \end{tabular}
    \end{adjustbox}
\end{table}

\begin{figure}[t]
    \centering
    \includegraphics[width=0.75\columnwidth]{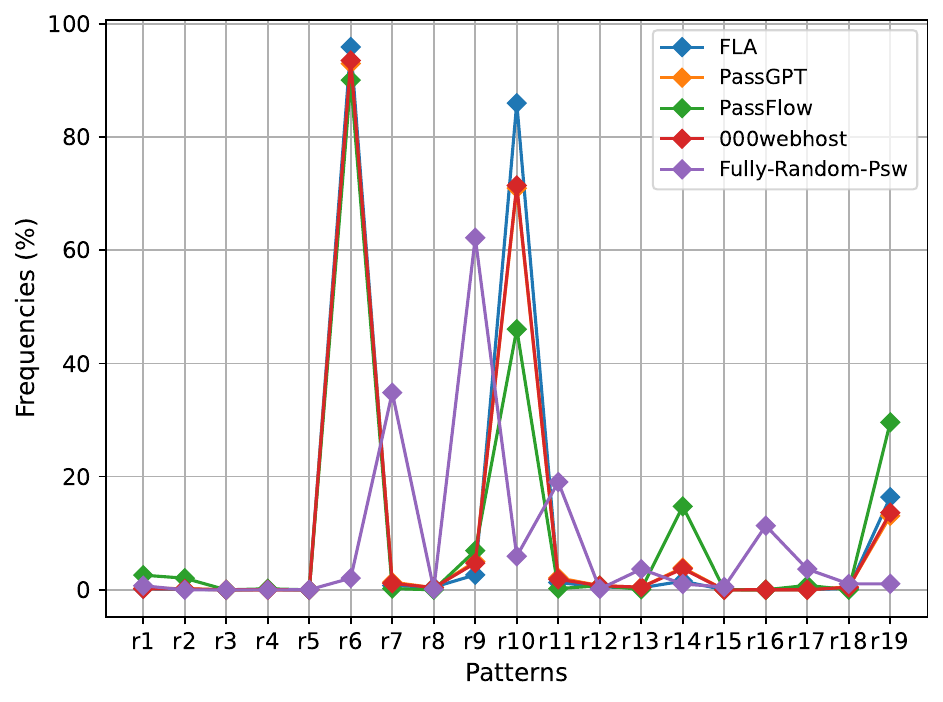}
    \caption{Distribution of 19 patterns in the 000webhost test dataset and in passwords generated by FLA, PassGPT, and PassFlow, all trained on 000webhost.}
    \label{fig:rq7-imd-pattern}
\end{figure}

\subsubsection*{\textbf{IMD Analysis}}
To better understand the rationale behind the high values obtained with IMD across several models, we repeat the metric analysis using $10^7$ generated passwords. The results, shown in Table~\ref{table:rq7-appendix}, differ significantly from those obtained with $5 \times 10^8$ generated passwords (Table~\ref{table:rq7}), particularly for PassFlow, PLR-GAN, and FLA, which exhibit notably higher IMD scores in the smaller sample. To investigate this discrepancy, we compare the length distributions of the $10^7$ and $5 \times 10^8$ generated samples. We observe that, in the smaller set, models such as PassFlow and PLR-GAN—based on GS and DPG techniques, respectively—exhibit a strong bias toward shorter passwords. This behavior stems from the tendency of such models to initially saturate the space of simpler (i.e., shorter) passwords before extending to more complex ones. A similar pattern is observed with FLA, which outputs the most probable passwords. Consequently, when fewer samples are generated, the length distribution exhibits a strong bias toward shorter lengths, as shorter passwords are more probable. These findings confirm our hypothesis that IMD is highly sensitive to length distribution differences: the greater the deviation of the generated passwords’ length distribution from the real one, the higher the distance measured by IMD.

We now focus on the 000webhost dataset, which yields a particularly high IMD score for PassFlow when evaluating $10^7$ generated passwords. Although this score exceeds that obtained from random passwords, it does not necessarily indicate that PassFlow’s outputs are random or even close to random. To demonstrate this, we compute the IMD metric between PassFlow’s generated passwords and random passwords. The results, shown in Table~\ref{tab:rq7-analysis1}, are compared against the two baselines and reveal that PassFlow’s passwords are farther from random passwords than from real ones. We further examined the effect of aligning the length distribution of the random passwords with that of PassFlow’s 000webhost-generated passwords. As reported in Table~\ref{tab:rq7-analysis1}, this adjustment results in a higher IMD score, thereby confirming that IMD is sensitive to differences in length distribution.

Lastly, we investigate the impact of various degrees of mode failure, such as mode dropping and mode invention, on the IMD metric. For this analysis, we used the 19 patterns listed in Table~\ref{tab:regex}. As shown in Figure~\ref{fig:rq7-imd-pattern}, the pattern distribution of random passwords deviates substantially from the others. Mode invention is observed in patterns r7, r9, r11, r13, r16, and r17, while mode dropping occurs in commonly observed patterns such as r6, r10, and r19. These findings suggest that the IMD metric does not adequately capture the varying degrees of mode failure and further support the hypothesis that IMD primarily outputs high values due to a model’s inability to replicate the real password length distribution. Interestingly, PassGPT performs particularly well in approximating the pattern distribution of 000webhost, with its pattern distribution closely aligning with that of the real passwords. Additionally, FLA outperforms PassFlow in capturing this distribution, with the latter struggling to replicate patterns r10, r14, and r19. These observations are consistent with the results obtained using the MTopDiv metric in Table~\ref{table:rq7}, which explicitly accounts for such mode-related discrepancies.

\begin{table}[t]
    \centering
    \caption{Distance between human- and generative model-created passwords when generating $10^7$ passwords.}    
    \renewcommand{\arraystretch}{0.2}
    \resizebox{1\columnwidth}{!}{
    \begin{tabular}{m{1.2cm}<{\centering} m{1.6cm} m{1.6cm} m{1.6cm} m{1.6cm} m{1.6cm} m{1.6cm} }
        \toprule
        \raisebox{-0.22em}{\textbf{Models}} & \raisebox{-0.22em}{\textbf{CNN Div}}  & \raisebox{-0.22em}{\textbf{$\alpha-$Precision}} & \raisebox{-0.22em}{\textbf{$\beta-$Recall}} & \raisebox{-0.22em}{\textbf{Auth}} & \raisebox{-0.22em}{\textbf{IMD}} & \raisebox{-0.22em}{\textbf{MTopDiv}} \\
        \hline
        \raisebox{-0.25em}{PassGAN} & \gradientbar{16} & \gradientbar{38} & \gradientbar{7} & \gradientbar{13} & \gradientbar{62} & \gradientbar{1}\\
        \raisebox{-0.25em}{PLR-GAN} & \gradientbar{9} & \gradientbar{-13} & \gradientbar{5} & \gradientbar{9} & \gradientbar{18} & \gradientbar{0}\\
        \raisebox{-0.25em}{PassFlow} & \gradientbar{69} & \gradientbar{31} & \gradientbar{59} & \gradientbar{35} & \gradientbar{500} & \gradientbar{52}\\
        \raisebox{-0.25em}{PassGPT} & \gradientbar{6} & \gradientbar{-9} & \gradientbar{4} & \gradientbar{4} & \gradientbar{0} & \gradientbar{0}\\
        \raisebox{-0.25em}{VGPT2} & \gradientbar{33} & \gradientbar{26} & \gradientbar{36} & \gradientbar{6} & \gradientbar{120} & \gradientbar{15}\\
        \raisebox{-0.25em}{FLA} & \gradientbar{24} & \gradientbar{39} & \gradientbar{3} & \gradientbar{48} & \gradientbar{347} & \gradientbar{2}\\
        \bottomrule
    \end{tabular}
    }
    \label{table:rq7-appendix}
\end{table}

\end{document}